\documentclass[%
reprint,
superscriptaddress,
amsmath,amssymb,
aps,
pra,
floatfix,
float
]{revtex4-2}
\usepackage{physics}
\usepackage{lipsum} 
\usepackage{blindtext}
\usepackage[utf8]{inputenc}
\usepackage[latin9]{luainputenc}
\usepackage[symbol]{footmisc}
\usepackage{amsmath}
\usepackage{amssymb}
\usepackage{amsthm}
\usepackage{setspace}
\usepackage{graphicx}
\usepackage{braket}
\usepackage{mathrsfs}
\usepackage{float}
\usepackage[english]{babel}
\usepackage[matrix,frame,arrow]{xypic}
\usepackage{lineno}\usepackage{amsfonts}
\usepackage[colorlinks = true,linkcolor = red,urlcolor  = blue,citecolor = blue,anchorcolor = blue]{hyperref}
\usepackage[english]{babel}
\usepackage{bm}
\usepackage [autostyle, english = american]{csquotes}
\MakeOuterQuote{"}
\usepackage{pifont}
\usepackage{blindtext}
\usepackage{ragged2e}

\begin{document}
\preprint{APS/123-QED}

\title{Supersolid Rotation in an Annular Bose-Einstein Condensate 
coupled to a Ring Cavity
}
\author{Gunjan Yadav}
\affiliation{Department of Physics, Indian Institute of Technology Guwahati, Guwahati 781039, Assam, India}
\author{Nilamoni Daloi}
\affiliation{School of Physics and Astronomy, Rochester Institute of Technology, 84 Lomb Memorial Drive, Rochester, New York 14623, USA}
\author{Pardeep Kumar}
\email{pardeep.kumar@mpl.mpg.de}
\affiliation{Max Planck Institute for the Science of Light, Staudtstraße 2, 91058 Erlangen, Germany}
\author{M. Bhattacharya}
\email{mxbsps@rit.edu}
\affiliation{School of Physics and Astronomy, Rochester Institute of Technology, 84 Lomb Memorial Drive, Rochester, New York 14623, USA}
\author{Tarak Nath Dey}
\email{tarak.dey@iitg.ac.in}
\affiliation{Department of Physics, Indian Institute of Technology Guwahati, Guwahati 781039, Assam, India}

\begin{abstract}
    We theoretically investigate an annularly confined Bose-Einstein Condensate (BEC) coupled to a four-mirror ring cavity supporting traveling-wave optical modes. Under symmetric driving by counter-propagating Laguerre-Gaussian beams carrying equal and opposite orbital angular momenta, the system realizes  supersolid phases coexisting with persistent superfluid circulation. Specifically, we obtain a supersolid state if we start with a BEC of winding number $L_p$ as well as supersolid packets with coherent superpositions of two different BEC $L_p$ values. Under asymmetric pumping, realized with Laguerre-Gaussian beams of different orbital angular momenta, chiral symmetry is broken in the system, resulting in asymmetric cavity field amplitudes, directional density modulations, and tunable rotational dynamics of the resulting supersolid lattice. This leads to rotating supersolid density structures for a single winding-number state, and rotating wave packets for an initial superposition of rotational eigenstates. Finally, we probe the presence of Goldstone and Higgs modes which can be observed using minimally destructive measurements of the cavity output spectrum. Our mean-field theory reveals interference-driven rotation without physical stirring, and distinguishes our work from prior static cavity supersolids. Our results establish the ring cavity annular BEC as a versatile platform for generating chiral quantum matter, implementing rotation-sensing devices and generating atomtronic circuits with supersolids.
\end{abstract}
\maketitle


\section{Introduction}
A defining feature of superfluidity is the presence of persistent currents \cite{Vilchynskyy_LTP_2013}, which reveal the ability of a quantum fluid to flow without dissipation for long durations \cite{PhysRevLett.128.150401, PhysRevX.12.041037}. In this context, Bose-Einstein condensates (BECs) \cite{PhysRevLett.75.3969} offer a uniquely versatile and tunable platform \cite{RevModPhys.73.307, 3564-qkv7}, enabling the precise control of superfluid behavior \cite{PhysRevLett.99.260401, polo2025persistent, RevModPhys.94.041001}. More specifically, BECs in multiply connected geometries, such as toroidal or annular traps \cite{RevModPhys.81.647}, provide topological protection to quantized circulation \cite{PhysRevA.86.013629,Das2012}, allowing access to the regimes of stable and even supersonic rotation \cite{PhysRevLett.124.025301,Pandey2019}.  Such ring-shaped condensates present a versatile platform for both investigating fundamental physics \cite{PhysRevX.8.021021,PhysRevA.91.013602,PhysRevLett.123.250402} and supporting emerging applications \cite{PhysRevLett.126.170402, PhysRevLett.113.135302, PhysRevLett.127.113601, PhysRevA.110.043512, kalita2023pump, pradhan2024ring, PhysRevA.82.043605, zd1d-39d7}.  

An intriguing extension arises when superfluidity and crystalline order appear simultaneously, as in a supersolid \cite{RevModPhys.84.759}, an exotic quantum phase that combines long-range spatial ordering with frictionless flow \cite{ PhysRevX.9.021012, PhysRevLett.120.123601}. In such a state, particles self-organize into a periodic structure while maintaining global phase coherence \cite{PhysRevLett.124.143602}, a counterintuitive combination first envisioned in helium \cite{kim2004probable} and more recently realized in ultracold atomic systems \cite{recati2023supersolidity, PhysRevX.9.021012}, including in BECs coupled to optical cavities \cite{leonard2017supersolid, nagy2008self}. This dual solid–fluid nature of supersolids offers promising applications in quantum technologies, including more stable quantum simulators \cite{su2023dipolar}, sensors \cite{PhysRevLett.122.190801, pelegri2018quantum}, and engineered quantum materials with novel transport properties \cite{mivehvar2021cavity}.

To date, cavity-based realizations of supersolids have largely been restricted to elongated trapping geometries \cite{PhysRevResearch.2.043318}, such as cigar-shaped BECs \cite{PhysRevA.70.023604, PhysRevX.9.021012}, in which translational symmetry is broken by the supersolid along a single spatial dimension \cite{leonard2017supersolid}. While these simply connected systems demonstrate the coexistence of superfluidity and density order, they lack rotational degrees of freedom in the BEC, thereby preventing the realization of topologically protected persistent currents, a key property of superfluids \cite{PhysRevLett.85.2228, PhysRevA.103.013313}. As a result, the dynamical fluid character of supersolids remains only partially explored \cite{PhysRevA.103.013313}. A further limitation arises from the cavity geometry itself, which determines the nature of the emergent order: linear cavities support standing-wave modes that explicitly break translational symmetry and impose a static lattice potential, whereas optical ring cavities sustain degenerate counter-propagating traveling-wave modes that preserve continuous translational symmetry \cite{shore1991quantum, PhysRevLett.120.123601, domokos2003mechanical}. In the latter case, any density modulation must emerge spontaneously from the atom-light dynamics \cite{RevModPhys.85.553}, providing direct access to genuine self-organization and symmetry breaking \cite{ostermann2015atomic}, which are key ingredients for realizing a fully dynamical supersolid \cite{PhysRevA.61.043405,  PhysRevA.102.063309, PhysRevE.105.054214}.

Motivated by these considerations, in this work, we overcome both constraints discussed above by confining a toroidally trapped BEC inside an optical ring cavity. In this setting, the ring BEC naturally supports quantized circulation characterized by an integer winding number $L_p$ \cite{PhysRevA.86.013629,PhysRevLett.99.260401, PhysRevLett.110.025301}, allowing rotational flow to coexist with spatial ordering and enabling a direct interplay between superfluid transport and crystalline structure. At the same time, the ring cavity supports Laguerre-Gaussian (LG) modes that couple light to atomic rotation \cite{cheng2017degenerate}. These modes carry orbital angular momentum (OAM) $\ell \hbar$ through a phase factor $e^{i\ell \phi}$ \cite{yang2022generation}, providing a controlled mechanism for transferring angular momentum to the atoms \cite{PhysRevA.106.L011304, poli2025synchronization, PhysRevLett.124.045702, 8n5y-fyh7}. In our configuration, the supersolid order is encoded in angular rather than linear density modulations, giving rise to crystalline patterns along the azimuthal direction of the ring \cite{PhysRevResearch.6.033116, PhysRevA.111.033304}. Crucially, these structures can rotate uniformly without external stirring, emerging from coherent interference between different angular momentum modes. This self-organized rotation provides a direct manifestation of the interplay between superfluidity and crystallinity, establishing a regime in which persistent flow and spatial order coexist on equal footing and enabling a genuinely dynamical realization of the supersolid phase.

Building on this framework, our main findings are twofold. We investigate the system under two distinct OAM pumping configurations: symmetric and asymmetric. Under symmetric OAM pumping, where the two LG pumps carry equal and opposite OAM, we demonstrate that for a condensate initially prepared in a single rotational eigenstate with winding number $L_p$ \cite{PhysRevLett.99.260401}, the system undergoes a self-organization transition into a supersolid phase exhibiting the coexistence of density modulation and global phase coherence \cite{ilzhofer2021phase}. This phase supports collective excitations in the form of gapless Goldstone modes and gapped Higgs modes \cite{goldstone1961field, PhysRev.127.965}, reflecting the underlying broken symmetries \cite{guo2019low, PhysRevResearch.6.L042056}. While, if the condensate is prepared in a coherent superposition of persistent current states \cite{PhysRevA.82.063623, PhysRevLett.97.170406, PhysRevLett.102.030405}, interaction with the cavity fields results in supersolid \textit{wave packets} with fine stripes distributed along the ring \cite{6d1g-671p}. The number of wave packets is controlled by the difference of the two winding numbers $L_{p_1}$ and $L_{p_2}$, and the periodicity is controlled by the winding number $\ell$ of the LG pump. These wave-packet supersolid states likewise exhibit both Goldstone and Higgs modes \cite{guo2019low}, with distinct spectral signatures tied to their spatial structure. Under asymmetric OAM pumping, where the applied LG fields carry unequal OAM, the cavity fields imprint a tunable chiral bias, enabling independent control over the rotation direction and angular velocity of the supersolid patterns \cite{PhysRevA.77.041601}. For condensates initialized in a single winding number eigenstate, this gives rise to rotating supersolid density modulations, whereas for condensates prepared in coherent superpositions of winding-number eigenstates, the system forms rotating \textit{wave packets}, providing a flexible route toward engineering nonequilibrium chiral quantum phases. Together, our results open the door for exploring rotating supersolids and nonequilibrium quantum phases that combine spatial ordering with coherent flow \cite{dong2025non}. Beyond their fundamental significance, these systems provide new opportunities for engineering chiral quantum matter \cite{PhysRevLett.121.030404}, investigating OAM resolved self-organization \cite{PhysRevLett.121.113204}, and realizing rotation-sensitive quantum devices based on supersolid order \cite{PhysRevA.110.033322}.

The paper is organized as follows: In Sec. \ref{SecII}, we introduce the model of the system, formulate the system Hamiltonian and present the dynamical equations. In Sec. \ref{SecIII}, we focus on symmetric OAM pumping, where we separately analyze two physically distinct initial conditions. The first case involves a single BEC rotational eigenstate, for which we provide a detailed treatment of the mode expansion, collective excitations, and cavity spectrum. The second case concerns the superposition of BEC rotational eigenstates, followed by the analysis of collective excitations and the cavity spectrum.  In Sec. \ref{SecIV}, we address asymmetric optical OAM pumping, with single winding number state and superposition of rotational eigenstates, respectively. Finally, we conclude in Sec. \ref{SecV} by summarizing the main findings and outlining potential applications of our results.


\section{Theoretical Model}
\label{SecII}

\begin{figure}[ht!]
\centering
    \includegraphics[width=0.875\linewidth]{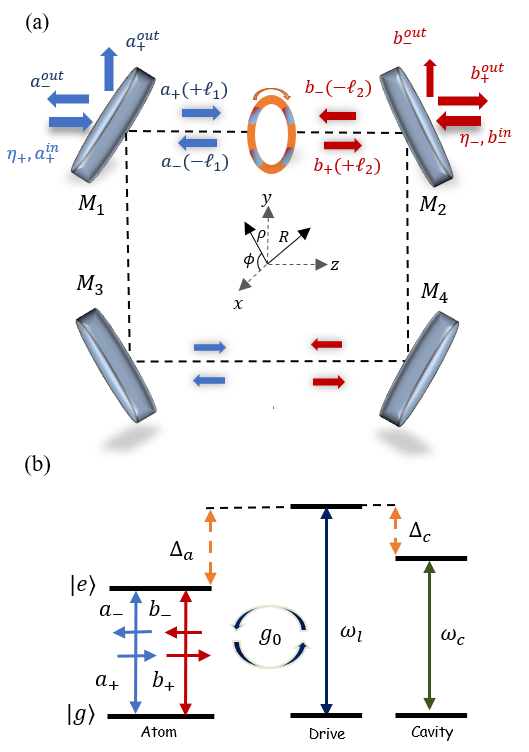}
    \caption{(a) Schematic of the model. A ring BEC coupled to co and counter-propagating cavity modes driven by light carrying OAM $+ \ell_1 \hbar$, and $- \ell_2 \hbar$. The input and output field through the cavity is denoted with $a^\text{in}_+$, $b^\text{in}_-$, and $a^\text{out}_+$, $a^\text{out}_-$, $b^\text{out}_+$, $b^\text{out}_-$, respectively.
    (b) Atom-cavity energy level structure. The atomic transition $|g\rangle \rightarrow |e\rangle$ is driven by a field of frequency $\omega_l$ with coupling strength $g_0$ and is detuned by $\Delta_a$. The cavity resonance frequency $\omega_c$ is red-detuned from the driving field by $\Delta_c$.}
    \label{fig:1}
\end{figure}

We consider a one-dimensional BEC of $^{23}\mathrm{Na}$ atoms confined in a ring trap of radius $R$, placed inside a four-mirror ring optical cavity aligned along the condensate symmetry axis, as shown in Fig.~\ref{fig:1}(a). In this configuration, the effective potential experienced by the atoms in the ring BEC enables the dynamics to be decoupled along the radial $(\rho)$, axial $(z)$, and azimuthal $(\phi)$ directions. Consequently, we employ a reduced one-dimensional description by considering only the dynamics along the azimuthal coordinate while assuming that the dynamics along the radial and axial degrees of freedom remain unchanged. Such a one-dimensional reduction is valid provided the total number of Na atoms satisfies the condition \cite{PhysRevA.74.023617},
\begin{equation}
    N < \frac{4R}{3a_{\text{Na}}}\Big(\frac{\pi\omega_\rho}{\omega_z} \Big)^{1/2}.
\end{equation}
Here $a_{\text{Na}}$ is the ground-state scattering
length of the Na atoms in the  condensate, $\omega_\rho$ and $\omega_z$ are the harmonic trapping frequencies
along the radial and axial directions, respectively. 

Owing to the four-mirror ring geometry, the optical cavity supports both co-propagating and counter-propagating traveling-wave LG modes with well-defined OAM, which form the natural eigenmodes of the ring cavity \cite{cheng2017degenerate}. This configuration is essential, as traveling-wave modes preserve directional information and enable efficient transfer of angular momentum to the ring condensate. The absence of standing-wave interference further eliminates fixed spatial nodes, making the system ideally suited for studying self-organization \cite{PhysRevLett.98.053603, PhysRevLett.124.143602}.

The cavity is coherently driven from two opposite directions $(\pm z)$ by LG laser beams with orthogonal polarizations, having pump strength $\eta_\pm$, frequency $\omega_{l}$, and OAM  $+ \ell_1 \hbar$, and $- \ell_2 \hbar$, respectively. These modes couple dispersively to the atoms of the ring BEC. The atoms forming the BEC are modelled as effective two-level systems, each with mass $m$, see  Fig.~\ref{fig:1}(b). The optical transition $\ket{g} \rightarrow \ket{e}$ corresponds to the sodium $D_1$ line ($3 ^{2}S_{1/2} \rightarrow 3 ^{2}P_{1/2}$) with a wavelength of approximately 
589.6 nm \cite{steck2000sodium}. This transition is coupled to the pump field with a single-photon coupling strength $g_0$. The pump fields are far detuned from the atomic transition frequency $\omega_{a}$, with a common atomic detuning $\Delta_a = \omega_l - \omega_a$. The atomic level structure and its coupling to the driving fields and cavity modes are illustrated in Fig.~\ref{fig:1}(b). Operating In the dispersive regime ($|\Delta_a| \gg g_0$), the excited state population stays negligible and responds much faster than the ground-state dynamics. Consequently, the excited state can be adiabatically eliminated, resulting in an effective description involving only the ground-state degrees of freedom.  \cite{RevModPhys.85.553} 

The two pump modes, denoted by $a_{+}$ and $b_{-}$, interact with the ring BEC and are coherently scattered into the counter-propagating cavity modes $a_{-}$ and $b_{+}$, respectively. This scattering process transfers a net angular momentum of $\pm 2\ell_1 \hbar$, and $\pm 2\ell_2 \hbar$ to atoms \cite{PhysRevA.106.L011304}. Both pump fields are detuned from the corresponding cavity resonance frequency $\omega_{c}$ by a fixed cavity detuning
$\Delta_c = \omega_l - \omega_{c}$.

\subsection{System Hamiltonian}
\label{secIIA}
Under the rotating wave approximation (RWA), the effective many-body Hamiltonian ($H_\text{eff}$) of the $N$ identical two-level atoms interacting with cavity fields is defined as a sum of the mean-field energy, cavity field energy, drive term, and interatomic interaction. The complete derivation of the Hamiltonian is provided in the Appendix. \ref{sec:appendixA}. Considering the two LG beams of OAM $+ \ell_1 \hbar$, and $- \ell_2 \hbar$, having orthogonal polarizations, the Hamiltonian can be written as
\begin{align}
    H_\mathrm{eff}
    = & \int_0^{2 \pi} \Psi^\dagger(\phi) \Big[-\frac{\hbar^2}{2I}\frac{\partial^2}{\partial \phi ^2} +  \hbar U_0\big(a_+^\dagger a_+ +  a_-^\dagger a_-  +  b_+^\dagger b_+\notag \\
    & +  b_-^\dagger b_- + a_+^\dagger a_- e^{-2 i \ell_1 \phi} +  a_-^\dagger a _+ e^{2 i \ell_1 \phi} + b_+^\dagger b_- e^{-2 i \ell_2 \phi}\notag\\
     &+  b_-^\dagger b _+ e^{2 i \ell_2 \phi} \big) \Big] \Psi(\phi) d\phi -\hbar \Delta_c [a_+^\dagger a_+ + a_-^\dagger a_- + b_+^\dagger b_+\notag\\
     & + b_-^\dagger b_- ] - i \hbar [ \eta_+ (a_+  - a_+^\dagger) + \eta_- (b_-  - b_-^\dagger)]\notag\\
      & + \frac{g}{2} \int_0^{2 \pi} \Psi^\dagger(\phi) \Psi^\dagger(\phi) \Psi(\phi) \Psi(\phi) d \phi.
     \label{Eq:1}
\end{align}
The first term of the integral in the square bracket represents the kinetic energy of the atoms, where $\hat{\Psi}(\phi)$ is the bosonic field operator at angular position $\phi$ on the ring, such that $[\Psi(\phi),\Psi^\dagger(\phi')] = \delta(\phi-\phi')$  and $I = m R^{2}$ is the atomic moment of inertia about the ring center. This term accounts for the rotational motion of atoms along the azimuthal coordinate $\phi$. The second term in the integral represents the potential energy, where $U_0 = g_0^2/\Delta_a$ denotes the effective atom–photon coupling strength, with $g_0$ being the single-photon atom–field coupling and $\Delta_a$  is the detuning of the pump field from the atomic transition. The first four terms in the potential energy part represent the AC Stark shifts experienced by the atoms due to photons in each of the four cavity modes 
These terms are spatially uniform and contribute an effective potential proportional to the intracavity photon number. The remaining four terms arise due to photon scattering from modes $a_\pm (b_\mp)$ to modes $a_\mp (b_\pm)$ and are associated with OAM transfers of $\pm 2\ell_1 \hbar$, and $\pm 2\ell_2 \hbar$ to the atom, which give rise to phase factors of $e^{\pm 2i \ell_1 \phi}$, and $e^{\pm 2i \ell_2 \phi}$ respectively \cite{PhysRevLett.124.143602}. Due to their orthogonal polarizations, the modes $a_\pm$ and $b_\mp$ do not interact directly; instead they couple via the ring BEC. The next term represents the energy of the cavity fields in the rotating frame of the pump laser $\omega_l$. The third term accounts for the coherent external driving of the cavity modes. The last integral term corresponds to interatomic interaction with strength $g = 2\hbar \omega_\rho a_\text{Na} / R$ \cite{PhysRevA.74.023617, PhysRevLett.123.195301}. In the following analysis, this interaction is neglected since it is typically much weaker than the cavity-mediated interactions and thus has a negligible effect on the system dynamics \cite{brennecke2013real}. 

\subsection{Equations of motion}
\label{secIIB}
The dynamics of the system are studied within the mean-field approximation, where each field operator is replaced by its expectation value $\langle a_\pm (t) \rangle = \alpha_\pm (t), \langle b_\pm (t) \rangle = \beta_\pm (t)$, and $\langle \hat{\Psi}(\phi, t) \rangle = \Psi(\phi, t)$. The coupled atom–cavity dynamics are described using the Heisenberg equations of motion. The cavity optical mode dynamics are governed by a set of equations \cite{PhysRevLett.124.143602, PhysRevLett.120.123601}
\begin{equation}
\frac{\partial }{\partial t} u(t) = M u(t) + v,
\label{Eq:2}
\end{equation}
where, $u(t) = \left(a_+(t), a_-(t), b_+(t), b_-(t)\right)^T$, $v = \left(\eta_+ , 0 , 0, \eta_-\right)^T$ and 
\begin{equation*}
    M = \begin{pmatrix}
    i\delta_c-\kappa_+ & -iU_0\mathcal{N}_1 & 0 & 0 \\
    -iU_0\mathcal{N}_1^* & i\delta_c-\kappa_+ & 0 & 0 \\
    0 & 0 & i\delta_c-\kappa_- & -iU_0\mathcal{N}_2 \\
    0 & 0 & -iU_0\mathcal{N}_2^* & i\delta_c-\kappa_-
\end{pmatrix}\;.
\end{equation*}
Here, $\delta_c = \Delta_c - N U_0$ denotes the effective cavity detuning, and $\kappa_{\pm}$ are the decay (loss) rates of the cavity modes. The atomic order parameters,
\begin{align}
\mathcal{N}_i = \int_0^{2\pi} \Psi^\dagger(\phi, t) e^{-2i\ell_i\phi} \Psi(\phi, t)d\phi\;,
\label{Eq:3}
\end{align}
quantifies the strength of the atomic density modulation along the ring, where $i \in \{1, 2\}$. The mean-field dynamics of the ring BEC coupled to a ring cavity is governed by: 
\begin{align}
     \frac{\partial}{\partial t} \Psi(\phi, t) = & \frac{-i}{\hbar} \Big[\frac{-\hbar^2}{2I}\frac{\partial^2}{\partial \phi ^2} +   \hbar U_0 \big(
    |\alpha_+|^2 + |\alpha_-|^2 + |\beta_+|^2\notag \\
    & + |\beta_-|^2 +  \mathcal{A}_1  +  \mathcal{A}_2 \big) \Big] \Psi(\phi, t),
    \label{Eq:4}
\end{align}
where $\mathcal{A}_1 = \alpha_+^* \alpha_- e^{-2 i \ell_1 \phi} + \alpha_-^* \alpha_+ e^{2 i \ell_1 \phi}$ and $\mathcal{A}_2 = \beta_+^* \beta_- e^{-2 i \ell_2 \phi} + \beta_-^* \beta_+ e^{2 i \ell_2 \phi}$. In this article, we will refer to Eq. (\ref{Eq:4}) as the Gross-Pitaevskii (GP) equation.  The condensate wavefunction is normalized to the total number of atoms $N$.

 \subsection{Numerical simulation details}
For a given initial state of a ring BEC, the coupled equations of motion Eqs.~(\ref{Eq:2}) and (\ref{Eq:4}) are solved numerically. The dynamical evolution of the cavity fields and condensate modes is obtained using real-time propagation based on the Runge–Kutta (RK45) method with controlled relative and absolute tolerances. The condensate wavefunction is discretized on a uniform grid of 256 points along the azimuthal coordinate \(\phi \in [0,2\pi]\), with periodic boundary conditions imposed to account for the ring geometry. Numerical convergence is carefully verified by increasing the spatial resolution and decreasing the integration time step, confirming that all relevant observables remain invariant within the desired numerical precision.

To determine the steady-state solutions, we employ a self-consistent imaginary-time propagation technique. In this approach, the system is evolved in imaginary time $\tau = i t$, which effectively projects the initial state onto the lowest-energy configuration compatible with the system constraints. During this procedure, the cavity field dynamics are treated within an adiabatic approximation, whereby the cavity fields are assumed to instantaneously follow the condensate dynamics~\cite{nagy2008self}. The imaginary-time evolution is initialized with the condensate wavefunction corresponding to a different set of winding number states, either a single winding number of superposition of winding number states.
As the evolution proceeds, the wavefunction relaxes toward a stationary solution. At each step of the propagation, the condensate wavefunction is explicitly normalized to preserve unit norm, ensuring numerical consistency. The convergence to steady state is established when the relative changes in both the wavefunction and the cavity field amplitudes fall below a predefined tolerance threshold.

The system parameters are chosen within a regime accessible to current cold-atom cavity experiments. We consider a ring-shaped atomic ensemble of radius $R = 12~\mu\mathrm{m}$, consisting of $N = 10^4$ atoms of mass $m = 23\,\mathrm{amu}$ corresponding to $^{23}$Na~\cite{metcalf1999laser,de2021versatile}. The radial confinement is provided by a harmonic trap with frequency $\omega_\rho / 2 \pi = 42~\text{Hz}$ \cite{de2021versatile}, and the interatomic interactions are characterized by an 
\textit{s}-wave scattering length $a_\text{Na} = 2.5~\text{nm}$ \cite{tiesinga1996spectroscopic}. The corresponding moment of inertia about the cavity axis is $I = mR^2$.
For symmetric pumping, we choose equal winding numbers for the driving fields, $+\ell_1 = +\ell$ and $-\ell_2 = -\ell$, such that the natural recoil frequency is defined as $\omega_r= {\hbar \ell^2}/{2I}$. Throughout the symmetric analysis, we set $\ell=10$, which yields $\omega_r \approx 9.6\times10^2~\mathrm{s}^{-1}\simeq 1~\mathrm{kHz}$, thereby setting the fundamental energy scale of the system.
The cavity decay rates are taken to be symmetric, $\kappa_+ = \kappa_- = \kappa = 50\,\omega_r \approx 50\,\mathrm{kHz}$, placing the system deep in the bad-cavity regime ($\kappa \gg \omega_r$), consistent with typical cavity setups \cite{PhysRevLett.112.115302} and enabling minimally destructive readout \cite{PhysRevLett.127.113601}. The atom-light dispersive coupling strength is chosen as $U_0 = -60\,\omega_r \approx -60\,\mathrm{kHz}$, corresponding to a red-detuned configuration where the pump frequency lies well below the atomic resonance.
The system is driven by two coherent pumps with equal amplitudes, $\eta_+ = \eta_- \equiv \eta$. Experimentally, the pump strength is related to the input laser power via $\eta = \sqrt{P_{\mathrm{in}} \kappa / (\hbar \omega_c)}$, where $\omega_c$ is the cavity resonance frequency \cite{PhysRevLett.127.113601}. Taking $\omega_c = \omega_a =  3.2 \times 10^{15}\,\mathrm{s}^{-1}$, we estimate that for $\eta$ varying from $0$ to $100\,\omega_r$, the corresponding input power lies in the range $P_{\mathrm{in}} \sim 0$ to $6.5 \times 10^{-14}\,\mathrm{W}$, which is well within experimentally achievable limits \cite{Brennecke_Science_2008}.

For asymmetric pumping, where $\ell_1\neq\ell_2$, we define the recoil frequency using the smaller OAM magnitude,
$\omega_r={\hbar \ell_{\min}^2}/{2I}, \ell_{\min}=\min(|\ell_1|,|\ell_2|)$. In our analysis, we consider the representative configurations $(\ell_1,\ell_2)=(8,4)$ and $(4,8)$, for which $\ell_{\min}=4$. This gives
$\omega_r= {\hbar(4)^2}/{2I}\approx 1.53\times10^2~\mathrm{s}^{-1}$. Using this recoil scale, the cavity detuning, dispersive coupling, and decay rates are chosen as
$\delta_c=-120 \omega_r, U_0=-60 \omega_r, \kappa_+=\kappa_-=50 \omega_r$ corresponding numerically to $\delta_c\approx -18.4~\mathrm{kHz}$, $U_0\approx -9.2~\mathrm{kHz}$, $\kappa_\pm\approx 7.7~\mathrm{kHz}$. For the same pump range $\eta\in[0,100\,\omega_r]$, the corresponding input power is estimated to vary from $0$ to $1.03\times10^{-14}~\mathrm{W}$.
\section{Symmetric OAM}
\label{SecIII}

To demonstrate the supersolid character of the ring BEC, we first consider the case of symmetric OAM pumping, in which two cavity modes are driven by optical beams carrying equal and opposite orbital angular momenta $\ell_1\hbar=\ell\hbar$ and $-\ell_2\hbar=-\ell\hbar$. 
This balanced transfer of angular momentum preserves the rotational symmetry of the effective cavity-induced potential and provides the simplest and most natural setting to investigate supersolid behavior in the ring geometry.


\subsection{Single winding number state}
\label{SecIIIA}
\subsubsection{Steady state response}

\begin{figure}[b]
\centering
    \includegraphics[width=\linewidth]{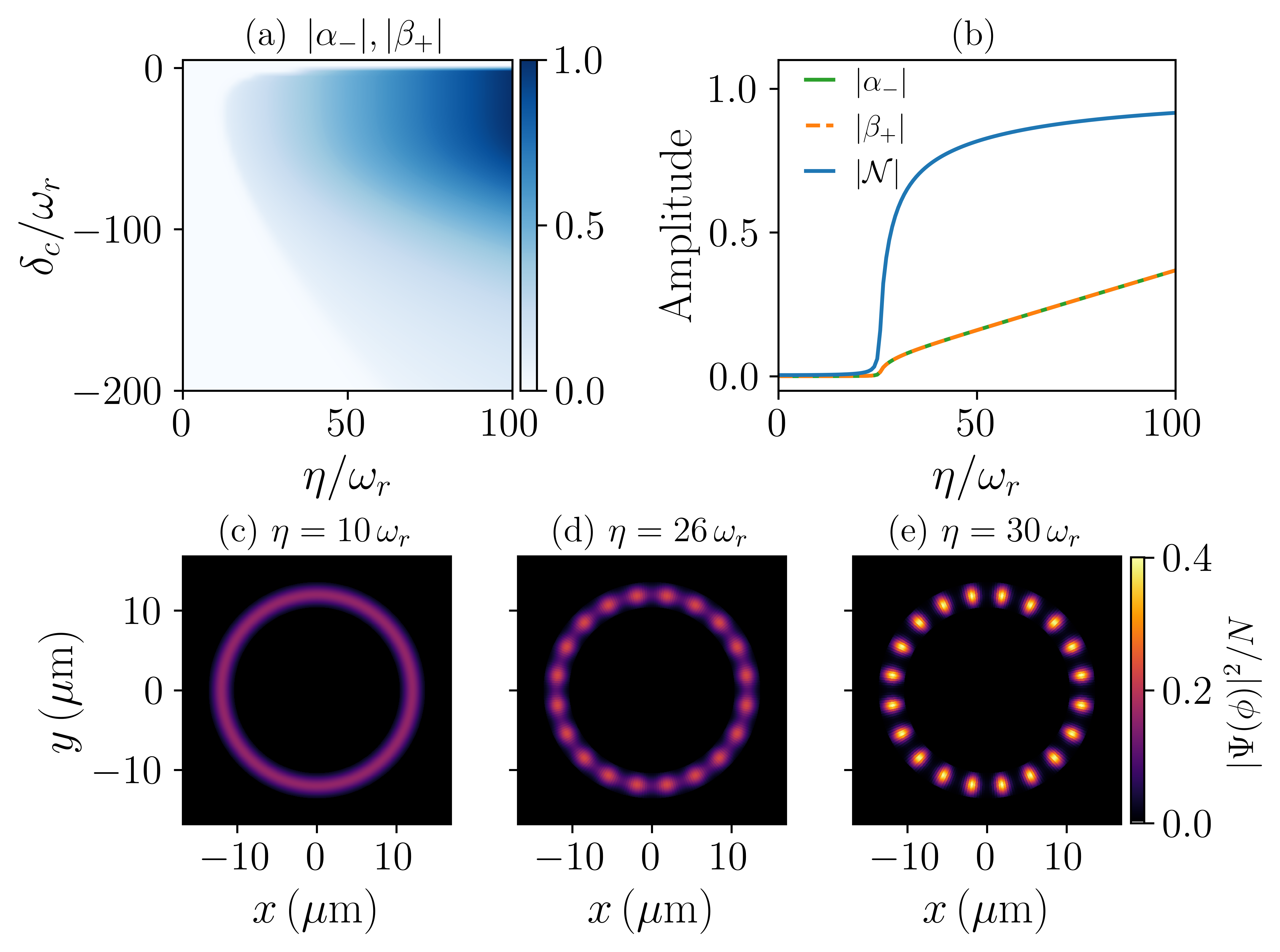}
    \caption{(a) Steady-state amplitudes of the scattered modes, $|\alpha_-|$ and $|\beta_+|$, as functions of the pump strength $\eta_\pm=\eta$ and the effective detuning $\delta_c$. The parameters used are $\kappa_+ = \kappa_- = 50\,\omega_r$, $N = 10^4$, $U_0 = -60\,\omega_r$, $\ell = 10$, and $L_p = 1$. Here $\omega_r = \hbar \ell^2/(2I)$ denotes the atomic recoil frequency, with $I = mR^2$, $m = 23\,\mathrm{amu}$, and $R = 12\,\mu\mathrm{m}$. (b) Steady-state behavior of the order parameter $\mathcal{N}$ (solid blue line) and the scattered-mode amplitudes $|\alpha_-|$ (solid green line) and $|\beta_+|$ (dashed orange line) as a function of $\eta/\omega_r$ for fixed detuning $\delta_c = -120\,\omega_r$. 
    (c)–(e) Density modulation of the condensate along the ring at different pump strengths: (c) $\eta = 10\,\omega_r$, (d) $\eta = 26\,\omega_r$, and (e) $\eta = 30\,\omega_r$.}
    \label{fig:2}
\end{figure}
\begin{figure}[t]
    \centering
    \includegraphics[width=\linewidth]{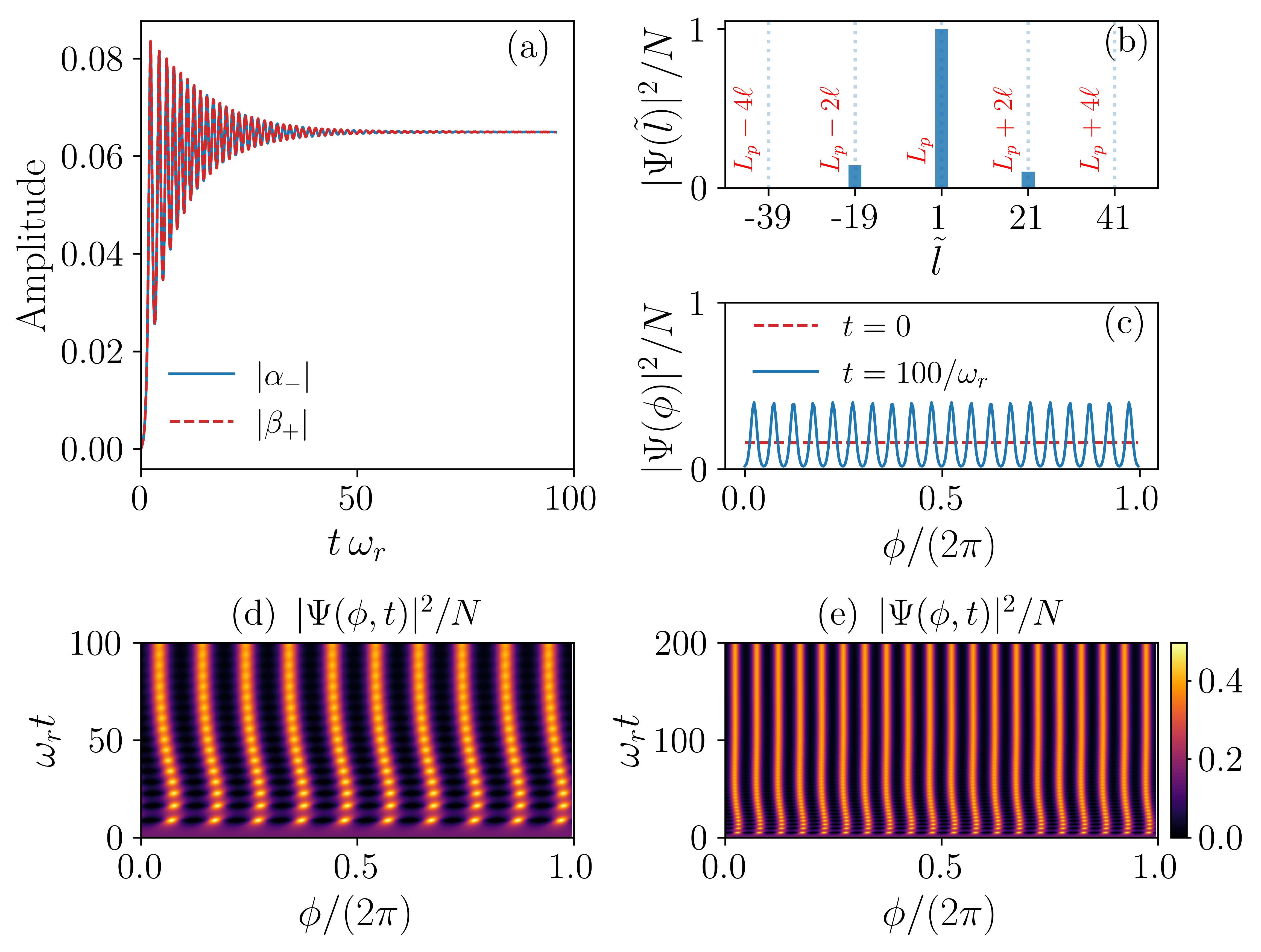}
    \caption{
    (a) Time evolution of the scattered field amplitudes $|\alpha_-|$ (blue) and $|\beta_+|$ (red).
    (b) Momentum-mode population $|\Psi(\tilde{l})|^2$.
    (c) Atomic density profile $|\Psi(\phi)|^2$ along the ring at $t=0$ (red dashed) and at $t=100/\omega_r$ (blue).
    (d,e) Spatiotemporal evolution of the atomic density $|\Psi(\phi,t)|^2$ along the ring for $L_p =1$ and (e) $\ell= 5$ (f) $\ell = 10$.}
    \label{fig:3}
\end{figure}
Here, we consider a ring BEC prepared in a single rotational state of winding number $L_p$ \cite{PhysRevA.86.013629,PhysRevLett.113.135302}. From the steady-state solutions, we extract the scattered-mode amplitudes $|\alpha_-|$ and $|\beta_+|$, as functions of the effective detuning $\delta_c$ and for symmetric pump strength, $\eta = \eta_\pm$, as shown in Fig.~\ref{fig:2}(a). This provides flexibility in tuning the system via detuning, which in turn modifies the critical pump threshold for self-organization. For the subsequent analysis of the steady-state response, we fix the detuning at $\delta_c = -120\omega_r$ \cite{PhysRevLett.112.115302}.
\begin{figure*}[t]
    \centering
    \includegraphics[width=1\linewidth]{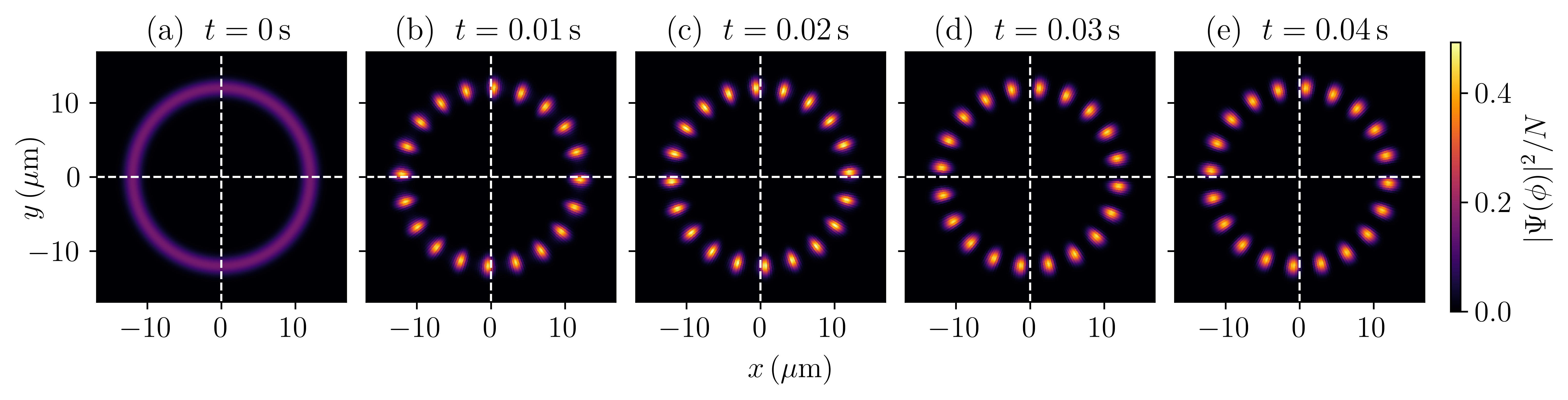}
    \caption{Snapshots of the density modulation along the ring at different evolution times for $\eta_+ = 30\,\omega_r$, $\eta_- = 35\,\omega_r$, and an initial winding number $L_p = 1$. For this parameter configuration, the supersolid pattern rotates in the direction of increasing azimuthal angle, i.e., from $0$ to $2\pi$, corresponding to an anticlockwise motion.}
    \label{fig:DM_t}
\end{figure*}
At fixed detuning, the response of the order parameter $\mathcal{N}$ and the scattered cavity fields as functions of the pump strength $\eta$, are shown in Fig.~\ref{fig:2}(b). With increasing $\eta$, both quantities remain negligible up to a critical pump strength $\eta_c \simeq 25\,\omega_r$, beyond which they exhibit a rapid increase. This threshold behaviour represents a phase transition from a homogeneous superfluid phase to an ordered supersolid phase driven by spontaneous $U(1)$ symmetry breaking and relative phase locking between the pump modes and their counter-propagating modes \cite{PhysRevLett.124.143602}, a process known as self-organization \cite{nagy2008self}. The supersolid phase consists of density modulation with periodic patterns along the ring, as shown in Figs.~\ref{fig:2}(c)-(e) for different $\eta$ values. Well below the threshold ($\eta = 10\,\omega_r < \eta_c$), the condensate density is nearly uniform, showing the homogeneous superfluid phase (absence of self-organization). As the pump strength approaches the critical strength ($\eta = 26\omega_r \approx \eta_c$), weak density modulations begin to emerge (buildup of self-organization). Above the pump strength ($\eta = 30\,\omega_r > \eta_c$), a periodic density modulation develops along the ring (emergence of stable supersolid phase). 

The self-organization mechanism observed in Fig. \ref{fig:2} can be explained by analyzing the real-time dynamics of the system, as shown in Fig.  \ref{fig:3}. Initially, the dynamics is dominated by a simple scattering process in which a condensate atom absorbs a photon from the pump field and re-emits a photon into the counter-propagating cavity mode. After an initial build-up, the oscillations of the cavity modes settle to finite steady values as shown in Fig. \ref{fig:3}(a). This absorption–emission cycle gives a momentum transfer to atoms of the condensate and transfers atoms from the initial winding number state $L_p$ to new winding number states $L_p + n\ell$, where $n = \{\pm 2, \pm 4....\}$. The resulting states form side modes of the original condensate, as shown in Fig. \ref{fig:3}(b) \cite{Brennecke_Science_2008, PhysRevLett.127.113601}. In this regime, only two side modes are significantly populated. The number of populated side modes can increase with the pump strength, as stronger driving enables higher-order momentum transfers. Once these side modes acquire a finite population, it interferes with the atoms in the initial momentum state of the condensate, generating a density modulation along the ring as shown in Fig. \ref{fig:3}(c).  
Initially (at $t=0$),
the density along the ring is homogeneous, while after sufficient time (at $t = 100/\omega_r$) it shows a periodic modulation, which confirms that the condensate is self-organized. Importantly, the emergent density lattice is not stationary; instead, it undergoes a continuous rotation around the ring. This rotational motion originates from the non-zero initial winding number $L_p$ of the BEC. 
The full spatiotemporal evolution of the density modulation is illustrated in Figs.~\ref{fig:3}(d) and \ref{fig:3}(e) for $\ell=5$ and $\ell=10$, respectively. In both cases, the system develops a periodic stripe pattern that remains stable over time. The angular periodicity is found to be $\pi/\ell$, showing that the lattice spacing is entirely controlled by the optical winding number. Consequently, increasing $\ell$ increases the number of density stripes around the ring while reducing their angular separation, leading to a finer supersolid lattice structure.

To reveal the dynamical nature of the emergent supersolid lattice, we introduce a slight imbalance in the pump strengths, choosing $\eta_+ = 30\,\omega_r$ and $\eta_- = 35\,\omega_r$, which amplifies the weak rotational motion of the ring condensate. Figs. \ref{fig:DM_t}(a-e) presents snapshots of the density modulation along the ring at different evolution times. The sequence clearly demonstrates that the emergent angular lattice undergoes a continuous rotation around the ring as time evolves. The direction and speed of this rotation are governed by both the initial condensate winding number $L_p$ and the OAM $\ell \hbar$ of the driving fields. In particular, reversing the sign of either $L_p$ or $\ell$ reverses the direction of rotation, providing direct control over the chirality of the supersolid lattice.
\subsubsection{Average angular momentum and Angular velocity}
\begin{figure}[t]
    \centering
    \includegraphics[width=\linewidth]{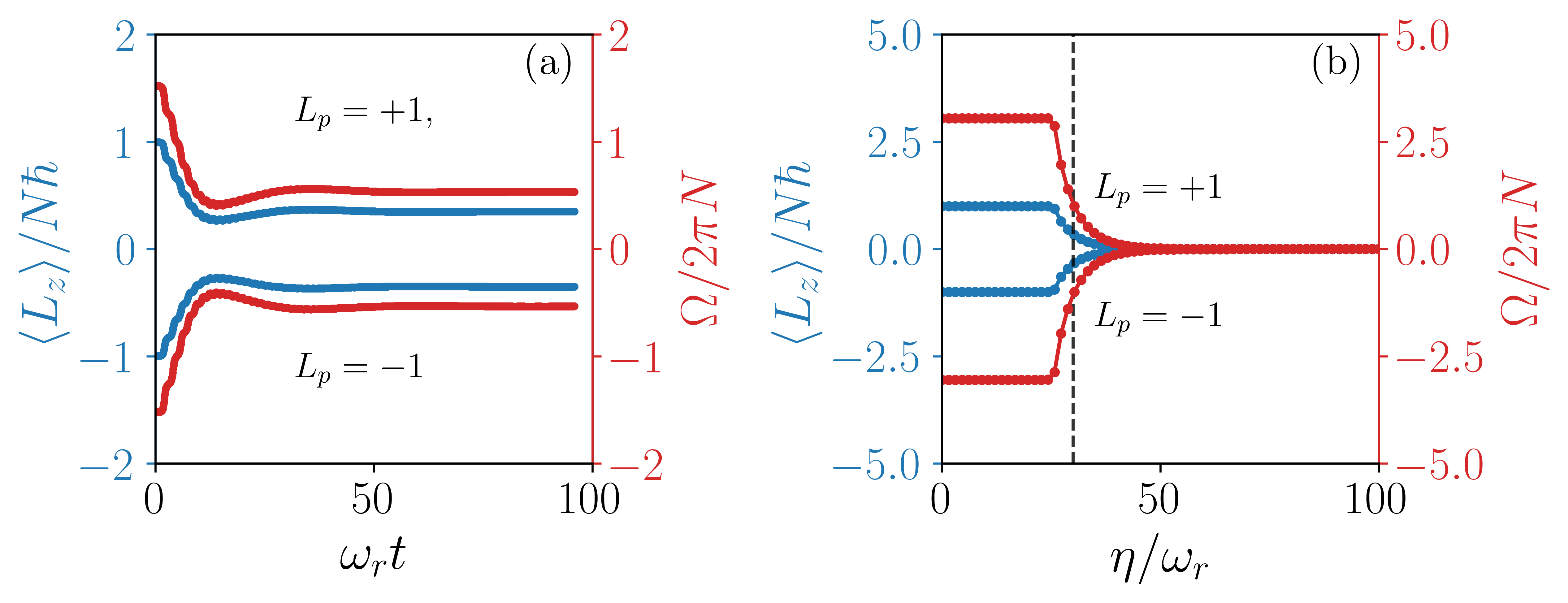}
    \caption{The average angular momentum $\langle L_z \rangle/N \hbar$ (blue, left axis) and the corresponding rotation velocity $\Omega/2\pi N$ (red, right axis): (a) Time evolution, and (b) Steady-state solution. The vertical dashed line in (b) indicates the critical pump strength near threshold, $\eta = 30 \omega_r$.}
    \label{fig:Lz}
\end{figure}

To quantify the rotational dynamics of the lattice, we next examine the time evolution of the average angular momentum $\langle L_z \rangle$ and the corresponding mean angular velocity, $\Omega=\langle L_z\rangle/{I}$,
as shown in Fig.~\ref{fig:Lz}(a). Both quantities evolve from their initial values and gradually relax toward finite steady-state values as time progresses \cite{PhysRevA.111.033304}. The emergence of a nonzero angular velocity directly confirms that the density-modulated supersolid rotates around the ring with an constant angular speed.

In the steady state, the condensate wavefunction can be expanded as, $\Psi(\phi, t) = \Psi(\phi) e^{-i \mu t/\hbar}$, where \(\mu\) is the chemical potential. 
The expectation value of the angular momentum for the condensate wavefunction $\Psi(\phi,t)$ at the steady state can be written as
\begin{equation}
    \langle L_z \rangle = \int_{0}^{2\pi}  \Psi^{\dagger}(\phi)\left(-i\hbar \frac{\partial}{\partial \phi}\right)\Psi(\phi) d\phi.
    \label{Eq:B1}
\end{equation}
Due to the nonlinear and self-consistent coupling between the cavity fields and the condensate, obtaining analytical expressions for $\Psi(\phi)$ and $L_z$ is not possible. The condensate evolution is governed by coupled equations of motion that depend explicitly on the intracavity field amplitudes, which themselves are influenced by the atomic density distribution. Therefore, to determine the steady-state angular momentum, we numerically solve the coupled dynamical equations for both the cavity fields and the condensate modes. Then the calculated value of $\Psi(\phi)$ is used in Eq. \eqref{Eq:B1} to numerically compute $\langle L_z\rangle$. 
The steady-state solution of $ \langle L_z \rangle $, and angular velocity $\Omega$, for $L_p = \pm1$ is plotted and shown in Fig. \ref{fig:Lz}(b). For pump strengths $\eta<\eta_c$, the system stays in the initial state and both $ \langle L_z \rangle / N\hbar$ and $\Omega/2\pi N$ remain constant. As $\eta > \eta_c$, atoms start scattering into higher angular-momentum states. Around the operating pump $\eta = 30\omega_r$, this scattering is slightly asymmetric, due to the different frequencies $\omega_{n}=\hbar (L_{p}\pm 2l)^2/2I$ of the various modes. It gives a finite angular momentum and indicates rotation. For $L_p = +1$, the value of $\langle L_z \rangle / N \hbar$ and $\Omega/2\pi N$ are positive and for $L_p = -1$, the values are negative, which represents the two different directions of rotation. For larger values of pump, the angular momentum gradually decreases and finally approaches zero. This behaviour can be understood by expressing the $\langle L_{z} \rangle$, in terms of the population of the momentum modes. Using the mode expansion Eq. \eqref{Eq:6} in Eq.~\eqref{Eq:B1}, $\langle L_{z} \rangle $ can be written as  
\begin{equation}
    \langle L_{z}\rangle = \hbar L_p N + \hbar \ell \sum_n n |c_n|^2,
\end{equation}
where, $n \in \{0, \pm2, \pm4...\}$.
This expression clearly shows that the average angular momentum can also be determined by the population distribution among the quantized angular momentum
states. The population distribution $|c_n|^2/N$ of different angular momentum modes are shown in Fig. \ref{fig:4}(a). Below $\eta_c$, the populations of side modes are zero  hence the contribution to $\langle L_z \rangle$ comes from the initial $L_p$ state only, $\langle L_z \rangle / N \hbar = L_p =1$.  As the pump strength increases above the $\eta_c$, the side modes population start to increase. From Fig. \ref{fig:4}(a), the population $c_{-2} > c_{+2}$, $c_{-4} > c_{+4}$, and so on. The net angular momentum
\begin{equation}
    \frac{\langle L_z \rangle}{N\hbar}  =  L_p - \ell \Big[2\Big(|c_{-2}|^2 - |c_{+2}|^2\Big) + .....\Big]\;,
\end{equation}
\noindent start to decrease from $1$. As the pump strength exceeded the $\eta_c$, near our operating $\eta = 30 \omega_r$, due to finite population difference, it gives finite values of $\langle L_z \rangle / N \hbar = 0.35$ and $\Omega/2\pi N = 0.9$. At large $\eta$, the population difference becomes very small but not zero; hence, the net angular momentum part due to population difference of various modes balances the angular momentum due to the initial $L_p$ state and $\langle L_z \rangle \rightarrow 0$.
%

\subsubsection{Mode expansion}
\label{SecIIIA1}
\begin{figure}[b]
    \centering
    \includegraphics[width=\linewidth]{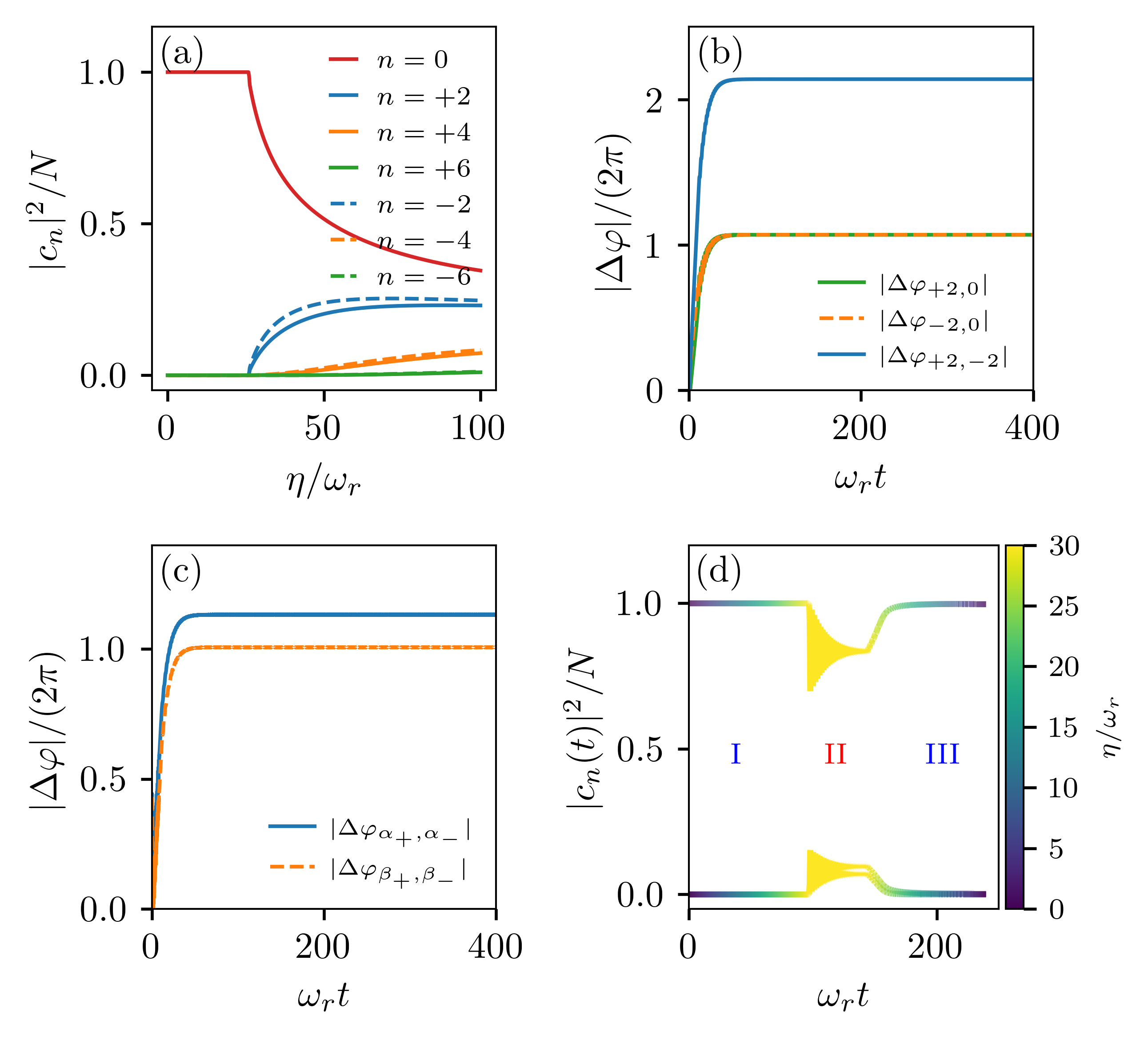}
    \caption{
    (a) Population $|c_n|^2$ of different atomic momentum modes $\{n=0, \pm2, \pm4, \pm6 \}$ of the BEC as a function of the normalized pump strength $\eta/\omega_r$.
    (c) Time evolution of the relative phase differences $\Delta\varphi/(2\pi)$ between the zero-momentum mode $n=0$ and the first-order excited momentum modes $n=\pm2$, near the threshold. (c) Time evolution of the relative phase differences $\Delta\varphi/(2\pi)$ between pump and scattered cavity modes.
    (d) Time evolution of mode populations $|c_n(t)|^2$, with $n =0 , \pm2$ during the pump ramp $(\eta/\omega_r)$ (colorbar), showing transitions between superfluid (I), supersolid (II), and superfluid (III) phases.}
    \label{fig:4}
\end{figure}
The redistribution of atoms into higher momentum states  [see Fig. \ref{fig:3}(b)] can be described by expanding the condensate wavefunction in angular momentum state basis \cite{PhysRevLett.127.113601}
\begin{equation}
\Psi(\phi,t) = \frac{1}{\sqrt{2\pi}} \sum_{n \in \mathbb{Z}} c_{n}(t)\, e^{i (L_p + n \ell) \phi},
\label{Eq:6}
\end{equation}
where $c_{n}(t)$ denotes the amplitude of the $n^\text{th}$ momentum mode and here $n \in \{0, \pm 2, \pm 4.....\}$. 
The order parameter defined in Eq. (\ref{Eq:3}) can then be expressed in terms of these momentum mode amplitudes $
\mathcal{N} = \sum_{n \in \mathbb{Z}} c_{n}^*(t)\, c_{n+2}(t)$, which quantifies the coherence between different momentum states of the condensate.
Substituting the mode expansion into the mean-field equations leads to dynamical equations for $c_{n}(t)$:
\begin{equation}
\frac{\partial}{\partial t} c_n(t) =  -i\left[ \omega_{n} \, c_{n} +  U_0 \left(\mathcal{A} \, c_{n+2} + \mathcal{A}^* \, c_{n-2} \right) \right],
\label{eq:7}
\end{equation}
where $\mathcal{A} = \alpha_+^* \alpha_- + \beta_+^* \beta_-$, and
\begin{equation}
    \omega_{n} = \frac{\hbar}{2 I} (L_p + n\ell)^2,
    \label{Eq:8}
\end{equation}
is the frequency corresponding to $n^\text{th}$ the momentum mode.
By solving these equations, the steady-state populations $|c_n|^2$ of the various atomic momentum modes $(n = \{0, \pm2, \pm4, \pm6 \})$ of the BEC as a function of the normalized pump strength is determined, and shown in Fig. \ref{fig:4}. Near the threshold, only the initial momentum mode \( n = 0 \) and the first-order side modes \( n = \pm 2 \) exhibit significant populations. The contributions from higher-order side modes \( (n = \pm 4, \pm 6) \) become appreciable only at increased pump strengths. In this regime, the essential physics of the system can be well captured within a three-mode approximation. 

Further, Figs. \ref{fig:4}(b) and \ref{fig:4}(c) show the time evolution of the normalized relative phase differences, $\Delta \varphi /(2\pi)$, for the condensate momentum modes $c_0$ and $c_{\pm2}$ and the cavity field modes, respectively. Following an initial transient regime, the relative phases gradually converge to constant values, indicating that the different momentum components establish a well-defined and time-independent phase relationship. An analogous behavior is observed for the cavity fields, where the pump modes and their corresponding scattered modes also evolve toward stable relative phases. This convergence to stationary phase differences signifies phase locking both among the relevant condensate modes and among the cavity field components. This stabilization of phases ensures a coherent coupling between the initial condensate and the side modes. 

Along with component-wise phase locking, the system also preserves global phase coherence across the participating momentum modes $c_0$ and $c_{\pm2}$, shown in Fig.~\ref{fig:4}(d). Varying the pumping from initially zero to the maximum value of $30 \omega_r$ and again reducing to zero, the system reversibly retraces the sequence of phases: superfluid \(\rightarrow\) supersolid \(\rightarrow\) superfluid \cite{PhysRevX.9.021012, ostermann2017probing}.
The absence of any observable hysteresis or delayed response during the forward and backward ramps indicates that the system remains dynamically stable throughout the evolution. This response provides evidence that the supersolid phase realized in the system is robust and stable within the considered parameter regime.

\begin{figure}[t]
    \centering
    \includegraphics[width=\linewidth]{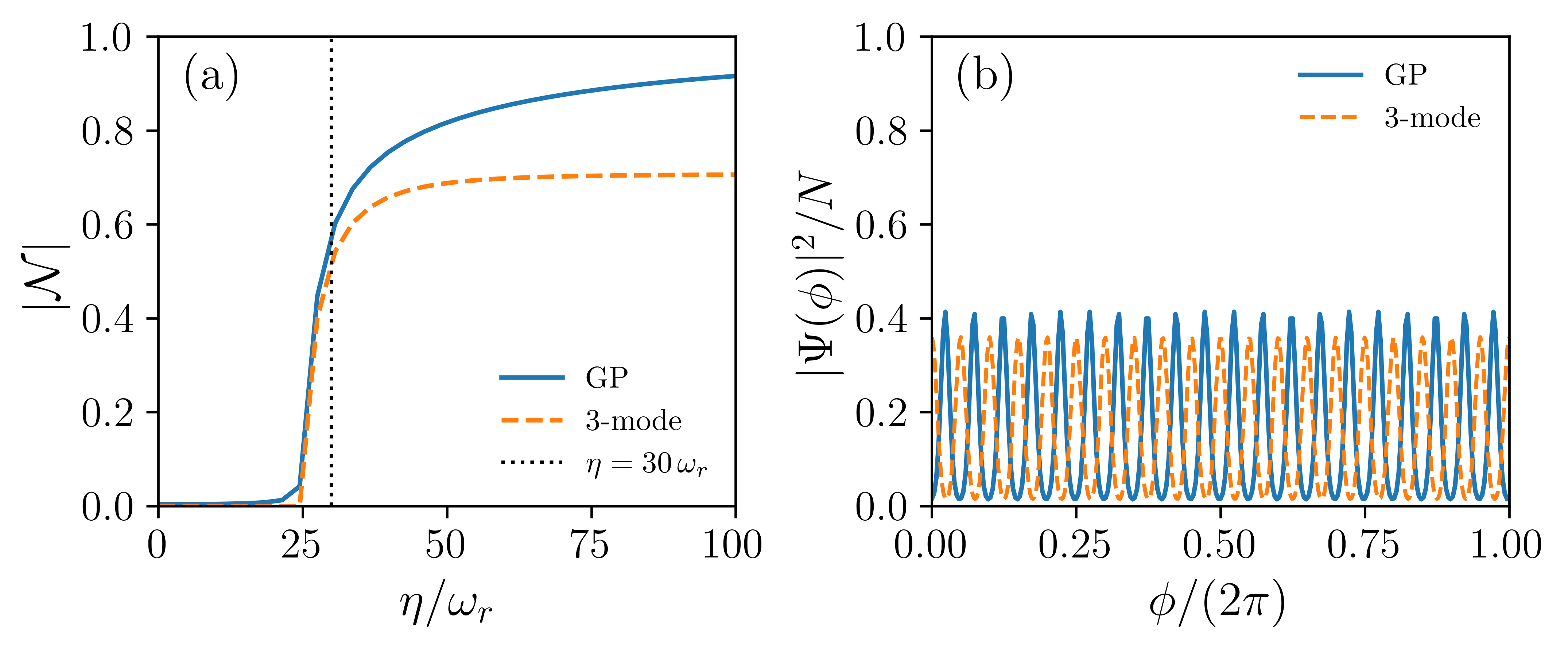}
    \caption{Comparison between the full GP equation results and the three-mode expansion model.
    (a) The ordered parameter as a function of the pump strength $\eta/\omega_r$. 
    (b) Density modulation $|\psi(\phi)|^2$ along the azimuthal direction of ring, at $\eta= 30 \omega_r$. 
    } 
    \label{fig:13}
\end{figure}
\begin{figure*}[htbp]
    \centering
    \includegraphics[width=\linewidth]{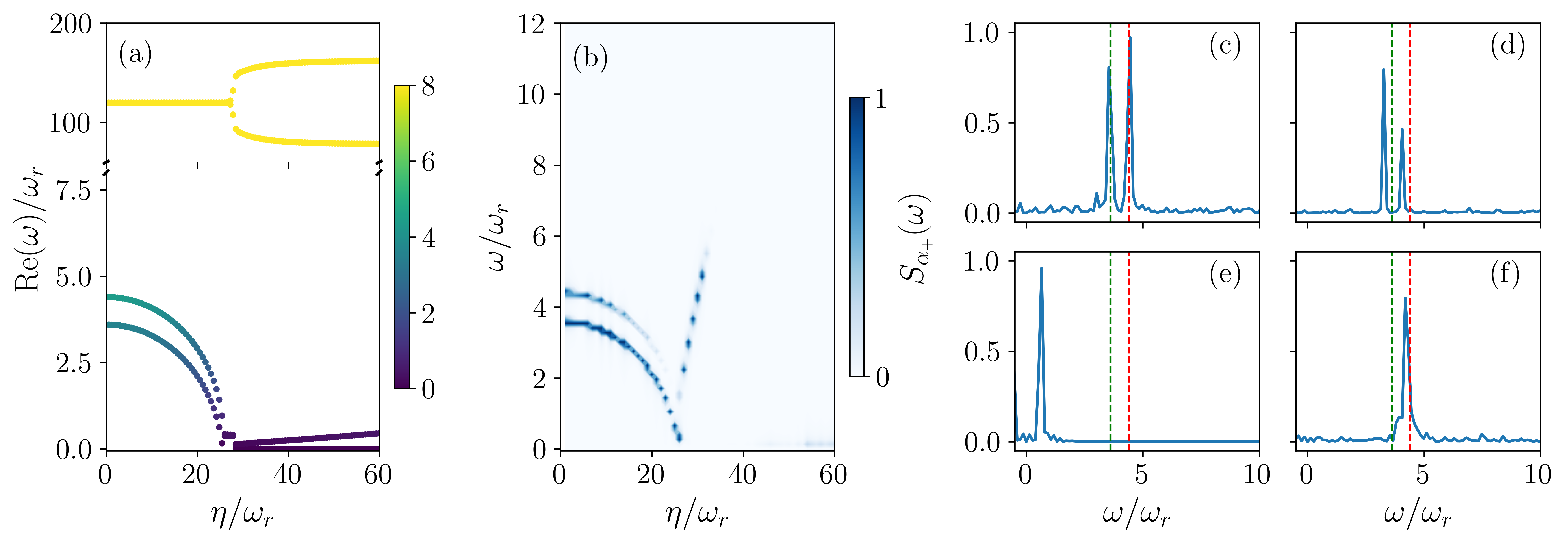}
   \caption{(a) Collective excitation spectrum obtained from the Bogoliubov analysis within the three-mode expansion. The two lower branches correspond to the BEC momentum modes $\omega_{\pm2}$, while the upper branch at $\omega = 120 \omega_r = -\delta_c$ corresponds to the cavity-photon excitation. 
    (b) Cavity output spectrum $S_{\alpha_+}(\omega)$ for different values of the normalized pump strength $\eta/\omega_r$. 
    (c)-(f) Cavity output spectrum at fixed pump strengths: (c) $\eta = 0.5\,\omega_r$, (d) $\eta = 10\,\omega_r$, (e) $\eta = 20\,\omega_r$, and (f) $\eta = 30\,\omega_r$. The vertical red and green dashed line corresponds to the analytical values of frequencies $\omega_{\pm2}$ as written in Eq. \ref{Eq:11}.}
    \label{fig:6}
\end{figure*}

Near the threshold, the three-mode expansion adequately captures the essential physics of the system. To verify this, we compare the results obtained from the full GP equation, Eqs. \eqref{Eq:2} and  \eqref{Eq:4}  with those mode expansions truncated up to three modes. In Fig. \ref{fig:13}, we exhibit that the two approaches show good agreement near the threshold. However, for a larger pump strength $\eta$, deviations appear. Fig.~\ref{fig:13}(a) presents the variation of the order parameter as a function of the pump strength, while Fig.~\ref{fig:13}(b) compares the corresponding density modulation along the ring at, $\eta = 30 \omega_r$. In particular, the qualitative behavior remains similar, but the amplitude predicted by the three-mode expansion decreases compared to the GP result. This occurs because, at higher $\eta$, additional higher-order momentum modes become populated. Therefore, for $\eta > 30\,\omega_r$, it becomes necessary to include modes beyond $\pm 2\ell\hbar$ in order to accurately describe the system dynamics. 
%

\subsubsection{Collective excitations}
\label{SecIIIA2}
The process of self-organization, for a single persistent current state, can be understood as a consequence of spontaneous \(U(1)\) symmetry breaking. 
The spontaneous breaking of continuous rotational symmetry leads to the appearance of low-energy collective excitations \cite{goldstone1961field}. In line with this expectation, we analyze the excitation spectrum of the system using the Bogoliubov method \cite{zhu2016bogoliubov, PhysRevLett.124.143602, PhysRevLett.120.123601}. Adding the fluctuation above the mean field solutions and then using the Bogoliubov transformation we form a matrix $\mathcal{M}_\text{BdG}$ satisfying (a detailed treatment is provided in Appendix \ref{sec:appendixB1b})
\begin{equation}
    \omega \mathbf{v} = \mathcal{M}_\text{BdG} \mathbf{v}
    \label{Eq:10}
\end{equation}
where $\mathbf{v} = (\delta \alpha_+^{(\pm)},
     \delta \alpha_-^{(\pm)}, 
     \delta \beta_+^{(\pm)}, 
     \delta \beta_-^{(\pm)},
     \delta c_{+2}^{(\pm)},
     \delta c_{-2}^{(\pm)})^T$. The eigenvalues $\omega$ of the Bogoliubov Eq. (\ref{Eq:10}) yield the collective excitation spectrum of the system. The real parts of the lowest-lying excitation frequencies, associated with both the condensate and the cavity photons, are shown in Fig.~\ref{fig:6}(a). For clarity, we display only the branch with positive real frequencies. In the weak-pump regime, $\eta << \eta_c$, the spectrum exhibits two dominant excitations at  $\omega_{\pm2}$, which correspond to the first-order momentum-sideband excitations of the condensate. These frequencies are given by
  \begin{equation}
        \omega_{\pm2} = \frac{\hbar}{2I}(L_p \pm 2\ell)^2
        =
        \begin{cases}
            \omega_{+2} = 4.41\omega_r \\
            \omega_{-2} = 3.61\omega_r.
        \end{cases}
        \label{Eq:11}
 \end{equation} 
In addition, a higher-frequency excitation appears at $\omega = 120 \omega_r = -\delta_c$ which corresponds to the cavity-photon mode. Together, these features characterize the elementary excitations of the coupled atom–cavity system in the normal phase. As the pump strength approaches the self-organization threshold $\eta_c$, the two lowest excitation frequencies progressively soften and move toward zero. Above the critical pump strength, the lower branch, associated with the mode $\omega_{-2}$, becomes soft and approaches zero frequency. This softening signals the emergence of a gapless Goldstone mode, which corresponds to phase oscillations of the supersolid order parameter \cite{doi:10.1126/science.aan2608}. Simultaneously, the higher branch at $\omega_{+2}$, remains finite and evolves into a gapped Higgs mode. This mode is associated with amplitude oscillations of the supersolid order parameter, and its frequency increases with increasing pump strength. The coexistence of the gapless Goldstone mode and the gapped Higgs mode constitutes a clear signature of supersolid behavior in the system.
 
\subsubsection{Cavity spectrum}
\label{SecIIIA3}
Signatures of the Goldstone and Higgs modes can be obtained from the cavity output spectrum of our proposed configuration. The spectrum is calculated using the input-output relation: $\mathcal{O}^\text{out}_\pm(\omega) = -\mathcal{O}^\text{in}_\pm(\omega) + \sqrt{2 \kappa}\mathcal{O}_\pm(\omega)$, where $\mathcal{O}_\pm \in \{\alpha, \beta \}$, and $\mathcal{O}_\pm(\omega)$ is a Fourier transform of $\mathcal{O}_\pm(t)$ \cite{RevModPhys.86.1391}. The term $\mathcal{O}^\text{out}_\pm(\omega)$ is the cavity output field, $\mathcal{O}^\text{in}_\pm(\omega)$ is input photon fluctuation, and $\mathcal{O}(\omega)$ is the solution of the cavity field by solving the GP equation in presence of stochastic noise.This noise is due to the optical fields as thermal noise is negligible at typical BEC temperatures \cite{PhysRevA.86.013629,PhysRevA.77.041601,Brennecke_Science_2008}. The phase quadrature of cavity transmission spectrum $S_{\mathcal{O}_\pm}(\omega)$ of operator $\mathcal{O}_\pm$ is defined by: 
\begin{equation}
    S_{\mathcal{O}_\pm}(\omega) = |\text{Im}(\mathcal{O}_\pm^\text{out}(\omega)|^2,
    \label{Eq:12}
\end{equation}
Fig. \ref{fig:6}(b) shows the cavity spectrum $S_{\alpha_+}(\omega)$ as a function of pump strength. To compare the cavity spectrum with the collective excitation, we first consider the case without fluctuations ($\mathcal{O}_\pm^\text{in}=0$). In this case the cavity output spectrum exhibits a behavior consistent with that obtained from the Bogoliubov analysis of collective excitations. Next, the spectrum is plotted in the presence of noise at different values of $\eta/\omega_r$, and the spectrum corresponding to $\alpha_+$ mode is shown in Figs. \ref{fig:6}(c)-(f). The spectrum corresponding to all four cavity modes is studied in the Appendix \ref{sec:appendixB1a}. At very low pump strength $(\eta = 0.5\omega_r < \eta_c)$, two distinct peaks appear in the spectrum, as shown in Fig.~\ref{fig:6}(c). These peaks are in excellent agreement with the analytically predicted frequencies $\omega_{\pm2}$ given in Eq. (\ref{Eq:11}). The corresponding positions of these modes are indicated by the green and red dotted lines in the figure. As the pump strength increases to $\eta = 10\,\omega_r$, the spectral peaks shift toward lower frequencies, as shown in Fig.~\ref{fig:6}(d). Upon approaching the critical pump strength $\eta_c$, the lower-frequency peak gradually disappears, as shown in Fig. \ref{fig:6}(e), showing the softening of the lowest excitation mode. For pump strengths well above the threshold (e.g., $\eta = 30\,\omega_r$), the peak position shifts to higher frequencies and moves away from zero, shown in Fig. \ref{fig:6}(f). The shift in peak shows the behaviour of Higgs modes.

\subsection{Superposition of two rotational eigenstates}
\label{SecIIIB}
Within symmetric OAM pumping, we now extend our analysis beyond the case of a single persistent-current state by considering an initial condensate prepared in a coherent superposition of two winding numbers \cite{PhysRevLett.97.170406, PhysRevA.77.041601,PhysRevLett.102.030405}, $L_{p_1}$ and $L_{p_2}$, written as \cite{pradhan2024ring}
\begin{equation}
    \Psi(\phi, 0) = \sqrt{\frac{N}{4\pi}}\left(e^{iL_{p1}\phi} + e^{iL_{p2}\phi}\right)\;.
    \label{Eq:13}
\end{equation}
Such states lead to interference in the angular degree of freedom, resulting in richer dynamical behavior compared to a single $L_p$ state. They are relevant for rotation sensing and provide a natural basis for quantum information encoding, while also enabling the study of mesoscopic quantum coherence in many-body systems \cite{PhysRevLett.97.170406, PhysRevA.82.063623}.

\subsubsection{Density modulation}
\begin{figure}[t]
    \centering
    \includegraphics[width=\linewidth]{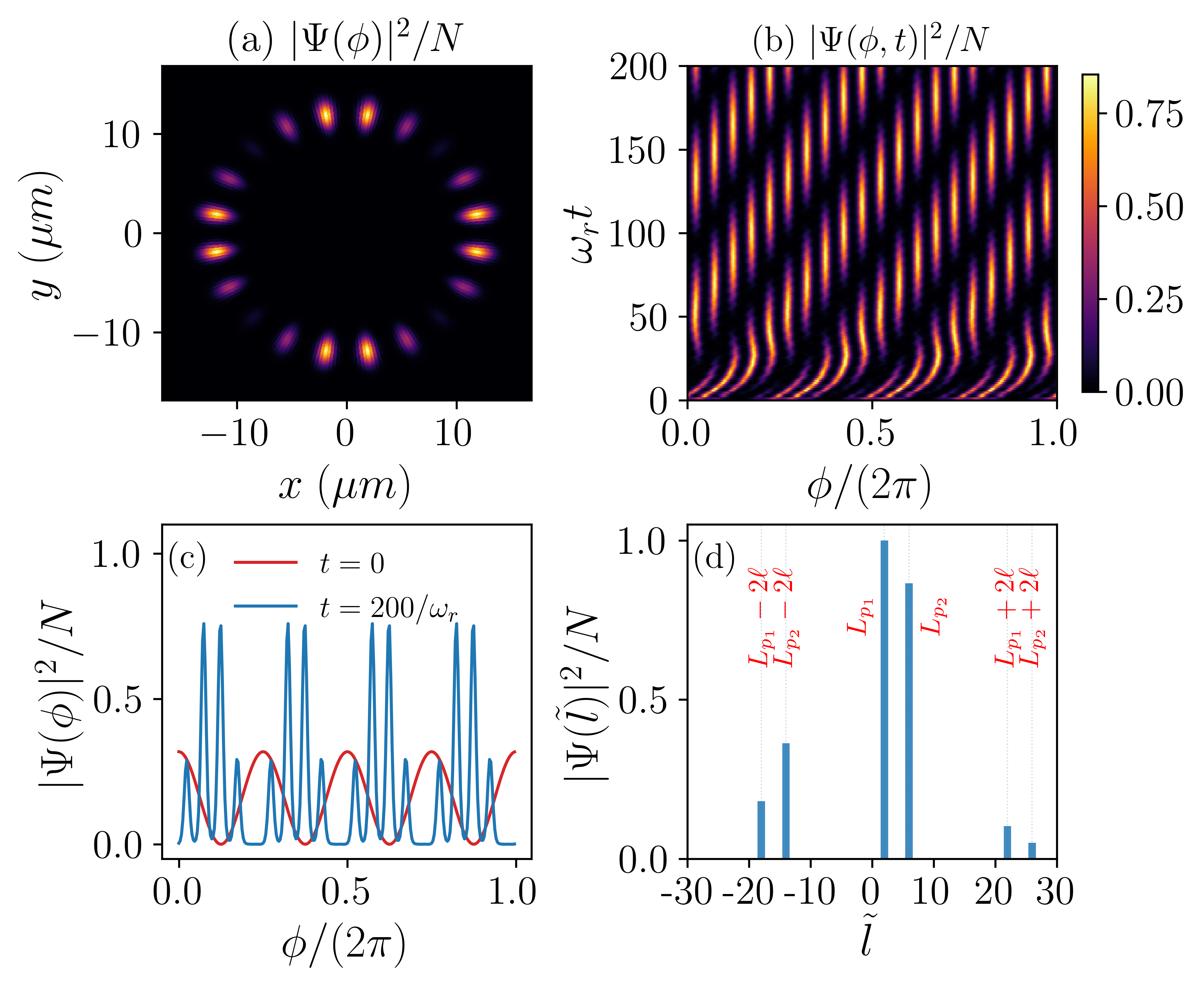}
    \caption{(a) The density modulation over the ring at $\eta = 30 \omega_r$. (b) The spatio-temporal evolution of density modulation.  (c) The density modulation with respect to angular position on the ring at initial time ($t=0$) and final time $(t = 200/\omega_r)$. (d) The Fourier transform of the BEC wavefunction showing the different momentum component excitation, and the rest of the parameters are the same as used in Fig. \ref{fig:2}.}
    \label{fig:7}
\end{figure}

\begin{figure*}[htbp]
    \centering
    \includegraphics[width=\linewidth]{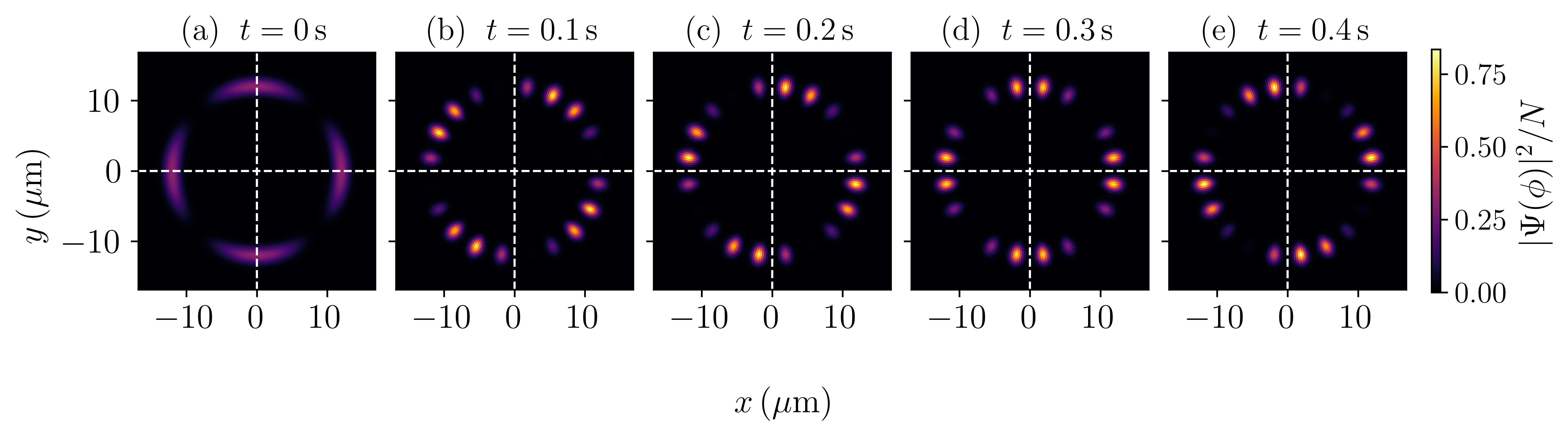}
    \caption{Snapshots of density modulation on the ring at different time intervals, for $L_{p_1} = 2$, $L_{p_2} =6$ and $\eta_\pm = 30 \omega_r$. The direction of rotation is counterclockwise.}
    \label{fig:DM_superpostion}
\end{figure*}

The superposition state expressed in Eq. \eqref{Eq:13} inherently supports interference between distinct angular-momentum components, giving rise to a weak but finite density modulation even in the absence of external driving. A representative example is shown in Fig.~\ref{fig:7} for $L_{p_1} = 2$ and $L_{p_2} =6$. As illustrated in \ref{fig:7}(a)-(c), the system develops four localized density envelopes rather than a uniform supersolid pattern along the ring. Within each envelope, finer density oscillations are present, forming stripe-like structures.
This behavior can be understood more generally. The interference between the two angular-momentum components fundamentally alters the spatial structure of the condensate. Unlike the single-$L_p$ configuration, which leads to a uniform supersolid pattern, the superposed state produces a modulated structure consisting of multiple localized density envelopes. Within each envelope, finer density oscillations are present, forming stripe-like patterns. The number of these envelopes is set by the winging number difference \(|L_{p_1}-L_{p_2}|\), while the finer stripes are distributed across the entire ring, with a total count of \(2\ell\). Consequently, the envelope periodicity is determined by $2 \pi/(L_{p_1} - L_{p_2})$. Additional examples for other combinations of $L_{p_1}$ and $L_{p_2}$ are presented in Appendix \ref{sec:appendixB2a}. These density peaks undergo a continuous rotation around the ring as time evolves, with an angular rotation frequency of $\Omega/2\pi = 1.25$, which is higher than that observed in the single $L_p$ case. Snapshots of the density modulation at different times are presented in Fig.~\ref{fig:DM_superpostion}, clearly illustrating the temporal shift of the density peaks along the ring.

From a momentum-space perspective, the dynamics involves excitations around each initial component, following $L_{p_1} \rightarrow L_{p_1} + n\ell$, and $L_{p_2} \rightarrow L_{p_2} + n\ell$, for $n \in \{\pm 2, \pm 4, \pm 6....\}$. The condensate wavefunction can therefore be expanded in this set of momentum modes. This structure is directly captured in the Fourier spectrum shown in Fig. \ref{fig:7}(d), where the excitation peaks corresponding to these processes are clearly visible.

\subsubsection{Average angular momentum and Angular velocity}
\begin{figure}[H]
    \centering
    \includegraphics[width=\linewidth]{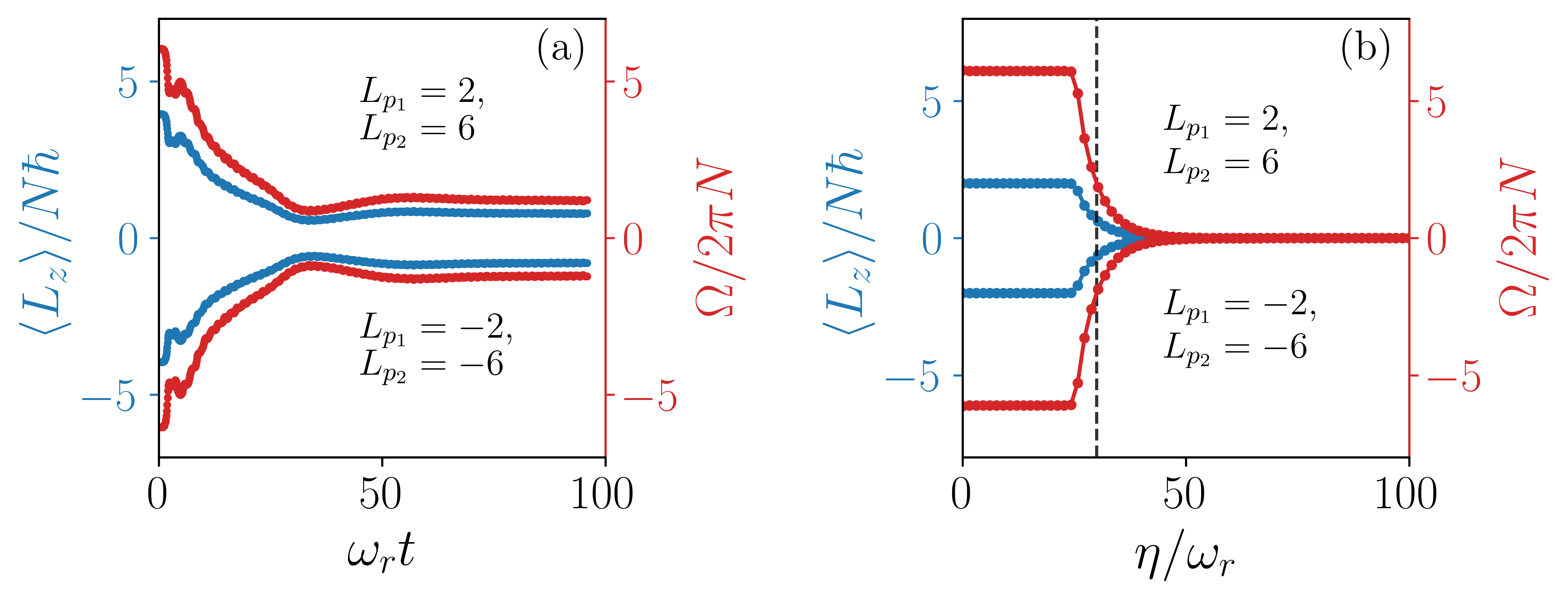}
    \caption{The average angular momentum $\langle L_z \rangle/N \hbar$ (blue, left axis) and the corresponding rotation velocity $\Omega/2\pi N$ (red, right axis). (a) Time evolution, and (b) Steady-state solution. The vertical dashed line in (b) indicates the critical pump strength near threshold, $\eta = 30 \omega_r$.}
    \label{fig:Lz_Lp1_Lp2}
\end{figure}
\begin{figure}[b]
    \centering
    \includegraphics[width=\linewidth]{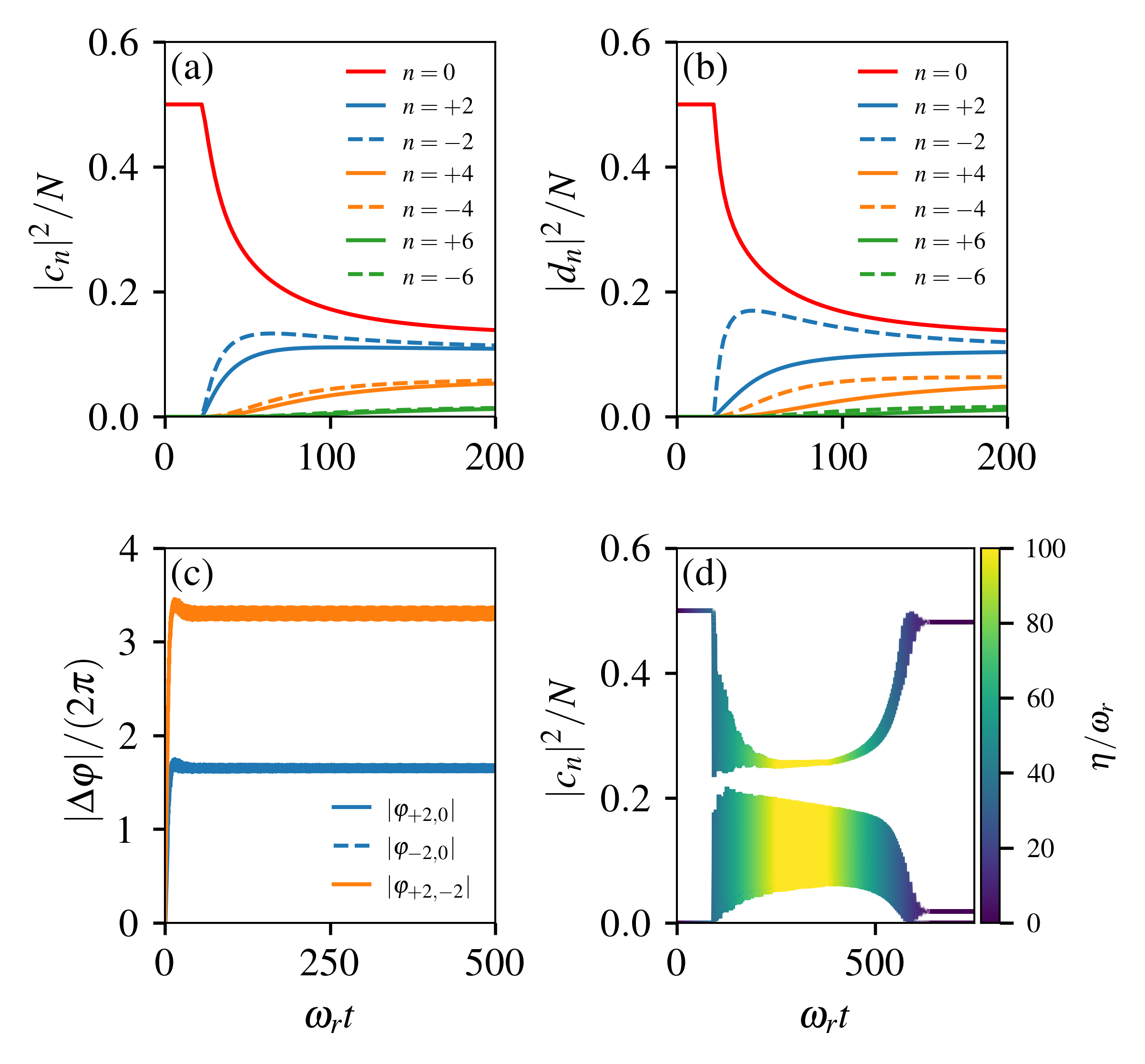}
    \caption{(a)-(b) The probability amplitude of different momentum mode excitations in the BEC corresponding to $L_{p_1}$ and $L_{p_2}$ states, respectively. (c) The time dynamics of phase difference $|\Delta \varphi|/(2 \pi)$ between different momentum modes. (d) The ramping of the momentum modes $c_0$, and $c_{\pm2}$ showing global phase coherence.}
    \label{fig:17}
\end{figure}
The time dynamics and steady-state behaviour of the average angular momentum and the corresponding angular velocity are presented in Figs.~\ref{fig:Lz_Lp1_Lp2}(a) and \ref{fig:Lz_Lp1_Lp2}(b), respectively. Similar to the single-$L_p$ case, the direction of rotation reverses upon changing the sign of the winding numbers $L_p$'s, confirming the controllable rotation direction of the rotating density-modulated state.
In the steady state, we obtain $
{\langle L_z\rangle}/{N\hbar}=0.5,
{\Omega}/{2\pi N}=1.8$,
demonstrating a faster and more robust persistent rotation of the supersolid lattice.
%
\subsubsection{Mode Expansion}
To gain further insight into the microscopic level, it is convenient to describe the condensate in the angular-momentum basis 
\begin{equation}
\Psi(\phi,t) = \frac{1}{\sqrt{4\pi}} \sum_{n \in \mathbb{Z}} \Big[ c_{n}(t) \, e^{i (L_{p_1} + n \ell)\phi} + d_{n}(t) \, e^{i (L_{p_2} + n \ell)\phi} \Big],
\label{eq:6}
\end{equation}
with amplitudes $c_{n}(t)$ and $d_{n}(t)$ corresponding to $n^{th}$ momentum mode, and here $n \in \{0, \pm2, \pm4...\}$. 
The atomic order parameter in this case becomes $\mathcal{N} = \sum_{n \in \mathbb{Z}^+} \left( c_{n}^* c_{n+2} + d_{n}^* d_{n+2} \right)$, and the coupled equations of motion are
\begin{subequations}
\begin{align}
\partial_t {c}_{n}(t) & =-i\Big[ \omega_{n}^{p1}\, c_{n} + U_0 \{ \mathcal{A} c_{n+2} + \mathcal{A}^* c_{n-2} \} \Big],\\
\partial_t {d}_{n}(t) & =-i\Big[ \omega_{n}^{p2} \, d_n +  U_0 \Big[ \mathcal{A} d_{n+2} + \mathcal{A}^* d_{n-2} \Big],
\end{align}
\label{eq:7}
\end{subequations}
where 
\begin{equation}
    \omega_{n}^{p_1} = \frac{\hbar}{2 I} (L_{p_1} + n\ell)^2, \quad \omega_{n}^{p_2} = \frac{\hbar}{2 I} (L_{p_2} + n\ell)^2,
    \label{Eq:C3}
\end{equation}
are the $n^\text{th}$ momentum mode frequencies corresponding to $L_{p_1}$ and $L_{p_2}$ mode excitations respectively and $\mathcal{A} = \alpha_+^* \alpha_- + \beta_+^* \beta_-$.
These equations describe the coherent evolution of the condensate under successive angular-momentum transfers due to the cavity photons.
The steady-state populations of the different momentum modes are shown in Figs. \ref{fig:17}(a) and \ref{fig:17}(b).
Near the threshold, only the initial momentum mode amplitude $c_0, d_0$ and first-order modes amplitudes $c_{\pm 2}, d_{\pm 2}$ acquire finite populations, while higher-order momentum modes $c_{\pm 4}, c_{\pm 6}, d_{\pm 4}, d_{\pm 6}$ remain negligibly populated. Hence, near the threshold, mode expansion can be truncated up to three modes. 

To see the phase relation between the different populated modes, we analyze the relative phase differences between $c_0$ and $c_{\pm 2}$. Fig.~\ref{fig:17}(c) shows that the phase differences acquire a constant values after a finite evolution time, which shows a phase locking among the different momentum components. To further confirm the global phase coherence of the supersolid state, we implement a pump-strength ramping protocol with $c_0$ and $c_{\pm 2}$. As illustrated in Fig.~\ref{fig:17}(d), the system undergoes a reversible transition from a superfluid to a supersolid phase and back to a superfluid as the pump strength is ramped up and reduced. The recovery of the initial phase-coherent superfluid state after the ramp cycle demonstrates the preservation of global phase coherence and highlights the reversible and coherent nature of the self-organization process for the superposition state.

\subsubsection{Analysis of collective excitations}
\label{SecIIIB1}
\begin{figure}[t]
    \centering
    \includegraphics[width=1\linewidth]{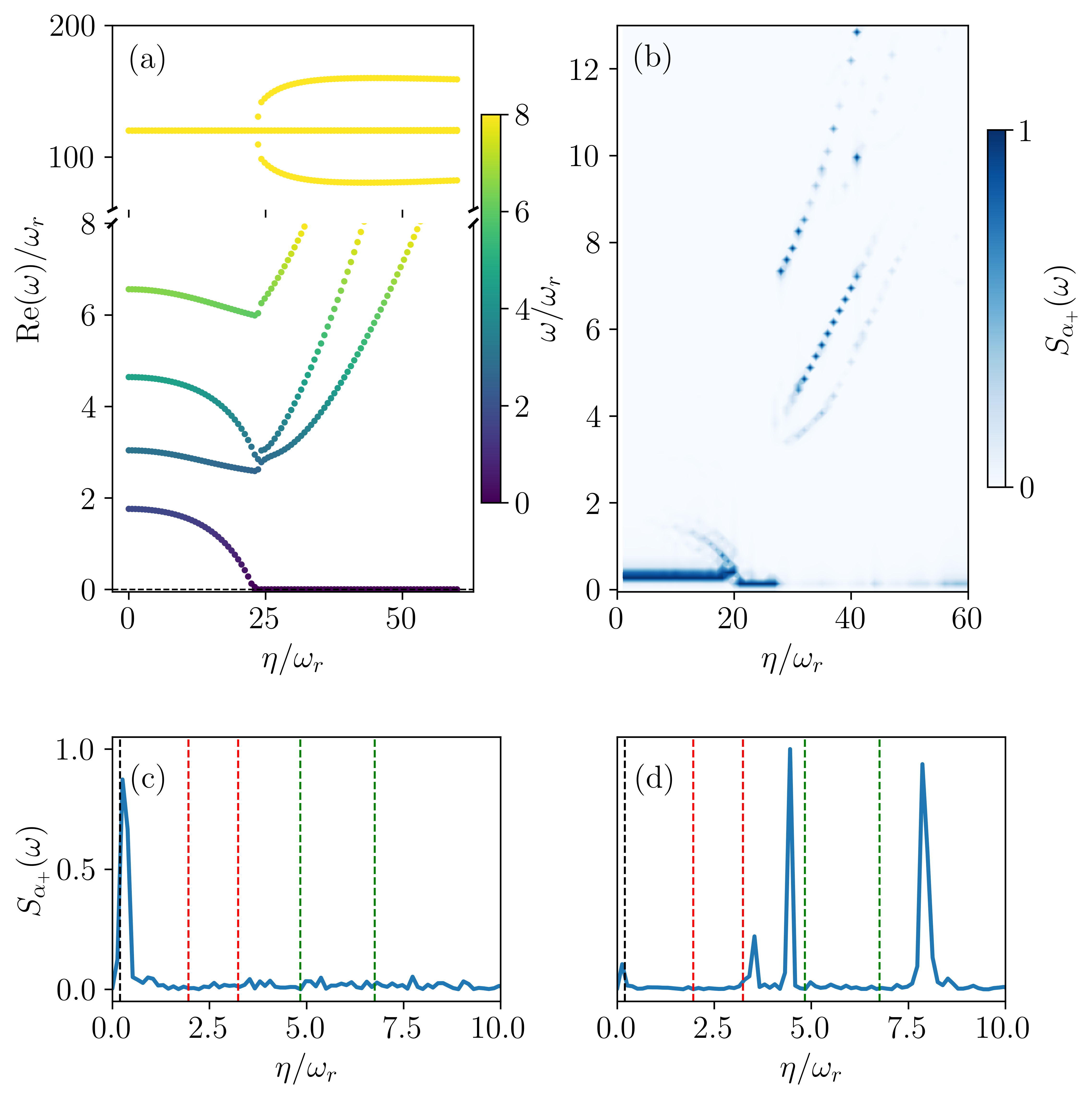}
    \caption{(a) Collective excitation spectrum obtained from the Bogoliubov analysis within the three-mode expansion. 
    (b) Cavity output spectrum $S_{\alpha_+}(\omega)$ as a function of $\eta/\omega_r$.
    (c),(d) Cavity spectra at fixed pump strengths: (c) $\eta = 0.5\,\omega_r$ and (d) $\eta = 30\,\omega_r$. 
    The green and red dotted lines in (c) and (d) denote the analytical values of the excitation frequencies, $\omega_{-2}^{p_1}, \omega_{-2}^{p_2}$ and $\omega_{+2}^{p_1}$, $\omega_{+2}^{p_2}$, respectively, as given in Eqs. (\ref{Eq:Lp1_freq}), and (\ref{Eq:Lp2_freq}).}
    \label{fig:8}
\end{figure}

We now turn to the collective excitation spectrum for a condensate prepared in a coherent superposition of rotational eigenstates. The detailed derivation is provided in Appendix. \ref{sec:appendixB2b}. The resulting spectrum, shown in Fig. \ref{fig:8}(a), exhibits four distinct Bogoliubov branches associated with the lowest excited angular-momentum components. These modes are located at
 \begin{equation}
        \omega_{\pm2}^{p_1} = \frac{\hbar}{2I}(L_{p_1} \pm 2\ell)^2
        =
        \begin{cases}
            \omega_{+2}^{p_1} = 4.84\,\omega_r, \\
            \omega_{-2}^{p_1} = 3.24\,\omega_r.
        \end{cases}
        \label{Eq:Lp1_freq}
 \end{equation}
 \begin{equation}
        \omega_{\pm2}^{p_2} = \frac{\hbar}{2I}(L_{p_2} \pm 2\ell)^2
        =
        \begin{cases}
            \omega_{+2}^{p_2} = 6.76\,\omega_r, \\
            \omega_{-2}^{p_2} = 1.96\,\omega_r.
        \end{cases}
        \label{Eq:Lp2_freq}
    \end{equation}
In addition, the cavity-field mode appears at $\omega = -\delta_c = 120\omega_r$.
As the pump strength approaches the critical threshold for self-organization, the lowest-lying excitation branch, $\omega_{-2}^{p_2}$, progressively softens and tends toward zero frequency. This behavior signals the emergence of a Goldstone mode associated with spontaneous symmetry breaking. In contrast, the remaining branches remain gapped and shift to higher frequencies with increasing pump strength, displaying the characteristic response of Higgs-like amplitude modes.
This behavior is directly reflected in the cavity-output spectrum shown in Fig.~\ref{fig:8}(b), which captures the same underlying excitation structure. In the weak-driving regime ($\eta < \eta_c$), the spectrum is dominated by interference between the two rotational components, giving rise to a single prominent peak at $(\omega_0^{p_1} + \omega_0^{p_2})/2 = 0.2\omega_r$. This feature is visible as the deep-blue intensity in Fig.~\ref{fig:8}(b) and is further highlighted by the black dotted line in Fig.~\ref{fig:8}(c). Upon crossing the threshold ($\eta > \eta_c$), the system enters the self-organized phase, where side-mode excitations become significant. As a result, multiple spectral peaks emerge, accompanied by a pronounced softening of the lowest excitation mode. This transition is clearly resolved in Figs.~\ref{fig:8}(c) and \ref{fig:8}(d), where three dominant peaks appear at $\omega_{+2}^{p_1}, \omega_{-2}^{p_1}$, and $\omega_{+2}^{p_2}$, in excellent agreement with the Bogoliubov spectrum shown in Fig.~\ref{fig:8}(a).

\section{Asymmetric OAM}
\label{SecIV}
We now move beyond the symmetric configuration and consider the general case in which the two pump fields carry unequal OAM, $\ell_1 \hbar$ and $-\ell_2 \hbar$, incident along the $\pm z$ directions. The condition $\ell_1\neq\ell_2$ explicitly breaks angular-momentum balance and imposes chirality on the system, giving rise to two competing ordering channels characterized by the order parameters $\mathcal{N}_1$ and $\mathcal{N}_2$. 

\subsection{Single winding number state}

\subsubsection{Steady-state response}
\begin{figure}[t]
    \centering
    \includegraphics[width= 0.95\linewidth]{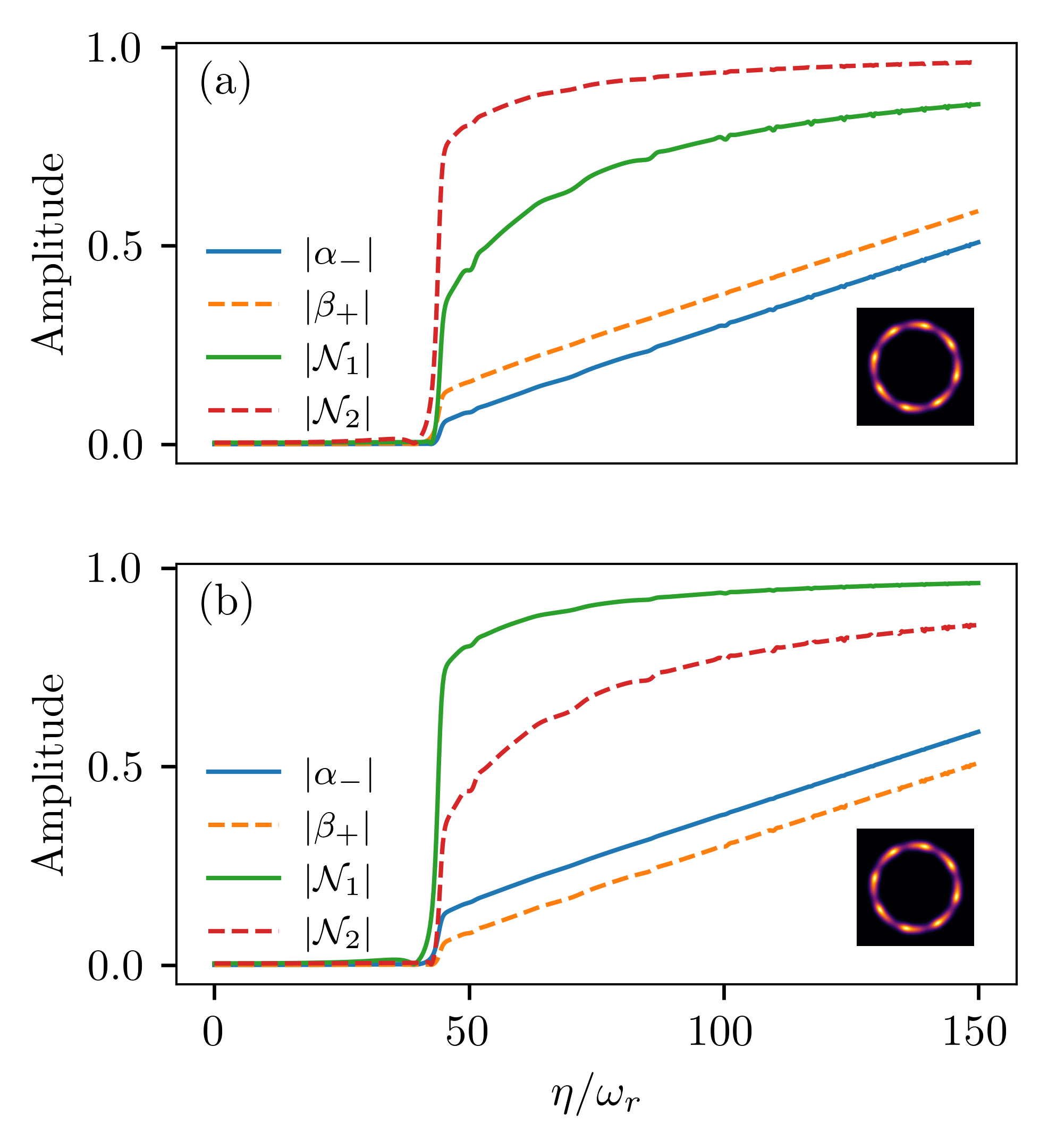}
    \caption{Steady-state response of the system for asymmetric OAM. (a) The variation of order parameter $\mathcal{N}$ and the scattered field amplitudes $|\alpha_-|$, $|\beta_+|$ as a function of pump strength $\eta/\omega_r$ for OAM index (a) $\ell_1 = 8$, $\ell_2 = 4$. (b) $\ell_1 = 4$, $\ell_2 = 8$. The inset shows the density modulation at $\eta = 100 \omega_r$.}
    \label{fig:9}
\end{figure}

Under asymmetric OAM pumping, we first consider the case where the ring BEC is initially prepared in a rotational eigenstate with a single winding number $L_p$. The steady-state behavior, shown in Figs. \ref{fig:9}(a) and \ref{fig:9}(b) reveals a pronounced asymmetry in both the cavity-mode amplitudes and the corresponding order parameters as functions of the pump strength $\eta/\omega_r$. We consider the representative cases $(\ell_1, \ell_2 ) = (8,4)$ and $(4,8)$. The mismatch in OAM enforces distinct azimuthal density modulations for the two scattering channels, which couple unequally to the cavity fields. This leads to unequal effective atom–light coupling and disparate growth rates, producing clearly separated steady-state amplitudes. The resulting imbalance is a consequence of the broken chiral symmetry. 

This asymmetry directly imprints on the real-space density distribution, shown in the insets at $\eta = 100\omega_r$, where the condensate preferentially selects a dominant ordering channel. This selection is governed by a competition between interaction-induced ordering and kinetic energy costs associated with different angular momentum modes. States with lower OAM are energetically favorable due to their reduced kinetic energy, while a larger order parameter enhances the gain from light–matter interaction (or self-organization). As a result, the system self-consistently selects the configuration that optimally balances these contributions, effectively minimizing the total energy. This leads to the emergence of a dominant mode characterized by both stronger ordering and lower angular momentum. Accordingly, for $\ell_1=8,\ell_2=4$, we find $\mathcal{N}_1<\mathcal{N}_2$ and the spatial structure is governed by the $\ell_2=4$ channel. Conversely, for $\ell_1=4,\ell_2=8$, we find opposite response $\mathcal{N}_1>\mathcal{N}_2$ and the $\ell_1=4$ channel dominates. In both cases, the emergent density pattern reflects the interplay between the order parameter and the OAM, with the dominant contribution associated with the smaller $\ell_{\min}=\min(|\ell_1|,|\ell_2|)$. Consequently, the number of maxima in the density modulation is governed by the smaller value of $\ell_{\min}$, yielding 8 maxima, as illustrated in the insets of Figs.~\ref{fig:9}(a) and \ref{fig:9}(b).
\begin{figure*}
    \centering
    \includegraphics[width=\linewidth]{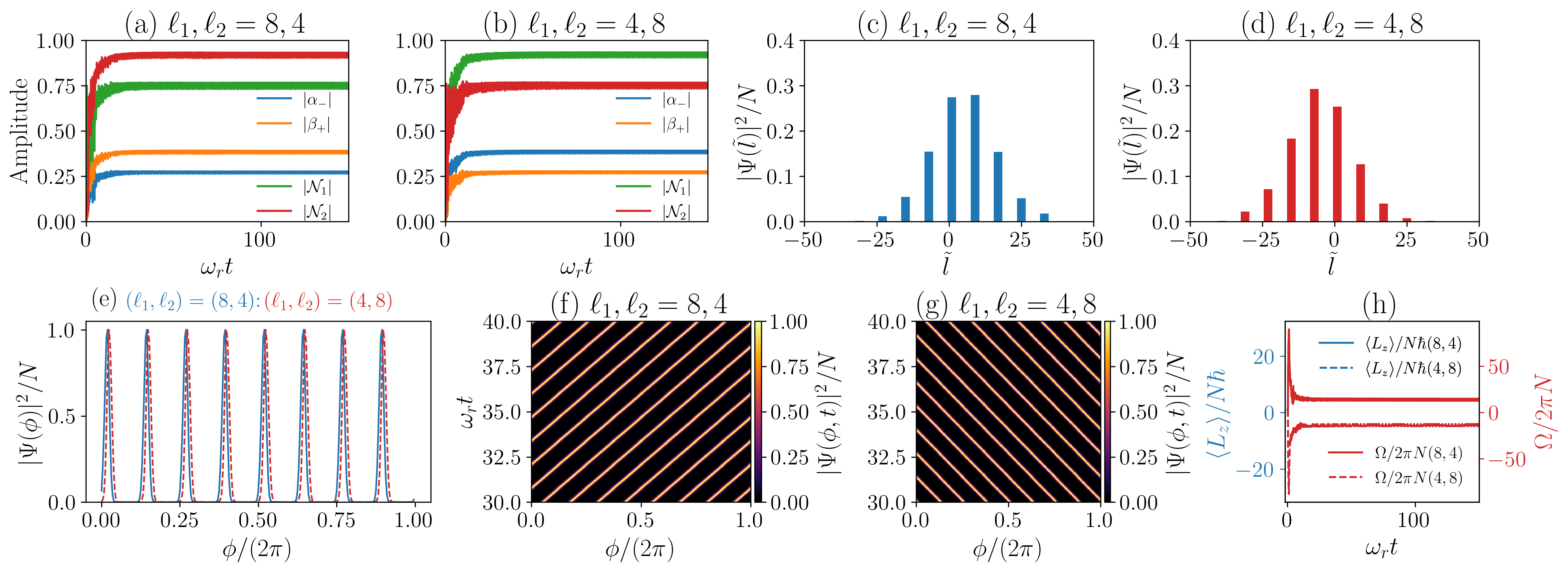}
    \caption{
(a,b) Time evolution of the scattered cavity-field amplitudes $|\alpha_-|$ and $|\beta_+|$, together with the order parameters $\mathcal{N}_1$ and $\mathcal{N}_2$, for $(\ell_1,\ell_2)=(8,4)$ and $(4,8)$, respectively.
(c,d) Momentum-space density (Fourier transform of the BEC wavefunction) $|\Psi(\tilde{\ell})|^2/N$ for $(\ell_1,\ell_2)=(8,4)$ and $(4,8)$, respectively.
(e) Azimuthal density modulation $|\Psi(\phi)|^2 / N$ for $(\ell_1,\ell_2)=(8,4)$ (blue) and $(4,8)$ (red).
(f,g) Spatiotemporal evolution of the condensate density for $(\ell_1,\ell_2)=(8,4)$ and $(4,8)$, respectively.
(h) Time evolution of the mean angular momentum $\langle L_z \rangle / \hbar N$ (blue, left axis) and the corresponding rotation frequency $\Omega/2\pi N$ (red, right axis).
}
\label{fig:10}
\end{figure*}

\begin{figure*}[htbp]
    \centering
    \includegraphics[width=\linewidth]{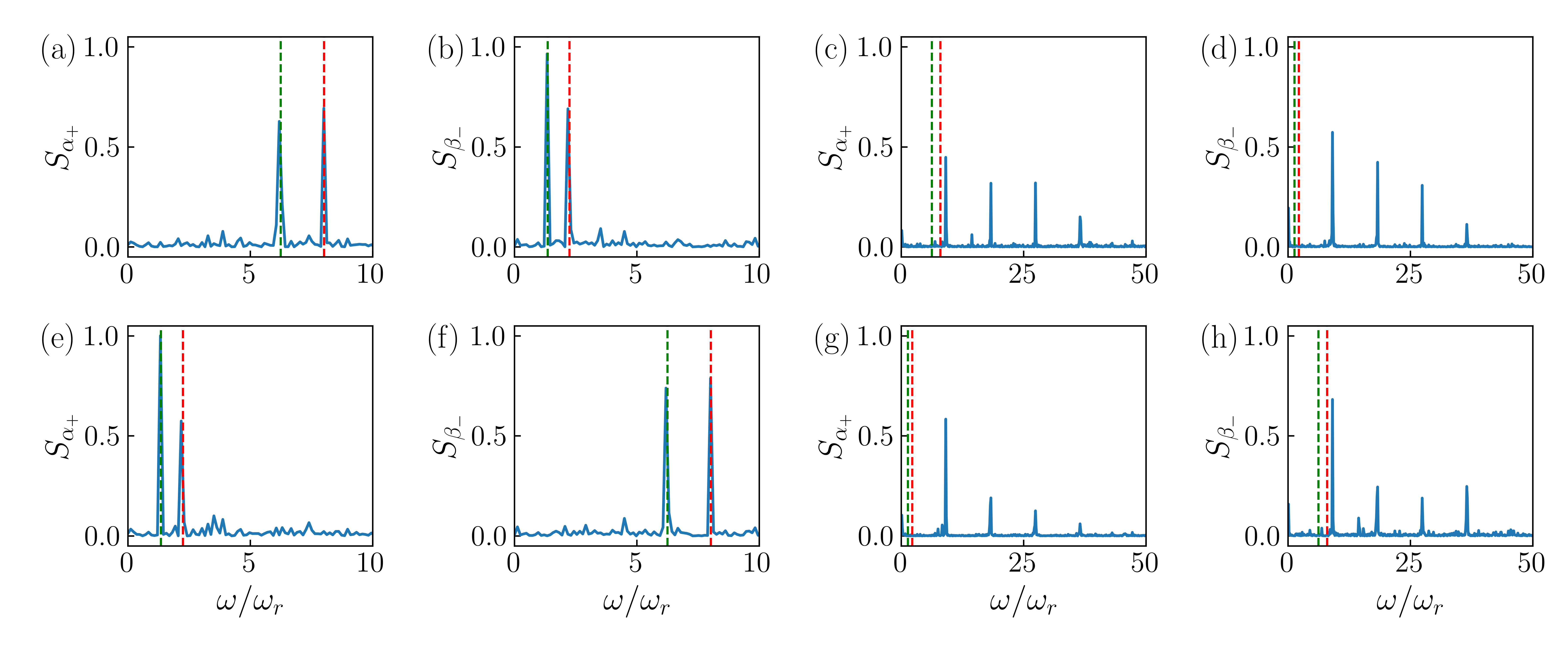}
    \caption{Cavity output spectrum for different combinations of OAM of the pump beams. (a–d) Results for $\ell_1 = 8$, $\ell_2 = 4$, and (e–h) results for $\ell_1 = 4$, $\ell_2 = 8$. Panels (a), (b), (e), and (f) are shown for pump strength $\eta = 0.5\omega_r$, whereas panels (c), (d), (g), and (h) are plotted for $\eta = 100\omega_r$. The green and red dotted lines indicate the analytical frequencies obtained from Eqs.~\ref{Eq:14} and \ref{Eq:15}, respectively.}
    \label{fig:11}
\end{figure*}
%

\subsubsection{Fourier analysis and density modulation}
In the immediate vicinity of the threshold, the system does not exhibit a stable response. This is due to the imbalance between the modes, which prevents efficient self-organization at low pump strengths. As a result, a sufficiently large driving strength is required to overcome this imbalance and stabilize the dynamics. In order to clearly capture the time evolution, we perform our analysis at a higher pump strength, choosing $\eta = 100\omega_r$. The time evolution of the scattered cavity fields and the corresponding order parameters is shown in Figs.~\ref{fig:10}(a)–(b) for the two representative configurations of $(\ell_1,\ell_2)$. After an initial buildup stage, the system evolves toward a steady state, and the amplitudes of the scattered cavity modes follow the same qualitative trends as observed in the steady-state analysis. 

The momentum-space distribution of the condensate, obtained from the Fourier transform of the BEC wavefunction, is displayed in Figs.~\ref{fig:10}(c) and (d). Distinct peaks appear at the winding number side modes $L_p \pm 2\ell_1$, $L_p \pm 4\ell_1$, and $L_p \pm 2\ell_2$, $L_p \pm 4\ell_2$. Since a relatively strong pump strength $\eta = 100\,\omega_r$ is considered, higher-order momentum modes are also populated in addition to the first-order side modes. Nevertheless, the probability amplitudes of the first-order modes remain significantly larger than those of the higher-order modes.

The resulting real-space density modulation, arising from the interference of these momentum components, is shown in Fig.~\ref{fig:10}(e) for $(\ell_1,\ell_2)=(8,4)$ (solid blue line) and $(4,8)$ (dashed red line). In both cases the spatial structure of the density modulation is identical; however, the direction of rotation is reversed. This behavior is clearly reflected in the time evolution of the mean angular momentum and the corresponding rotation frequency presented in Figs.~\ref{fig:10}(f)–(h). The opposite signs of $\langle L_z \rangle$ and $\Omega$ indicate that reversing the relative ordering of $\ell_1$ and $\ell_2$ leads to a reversal of the rotation direction of the supersolid lattice. At steady state, the calculated value of $\langle L_z \rangle / N \hbar = 5$, and $\Omega / 2 \pi N = 15$ for $\ell_1 = 8, \ell_2 = 4$, which is much greater compared to symmetric OAM.
%

\subsubsection{Cavity spectrum}
\label{SecIIIC1}
Due to the excitation of multiple higher-order momentum modes in this regime, a simple analytical description based on a truncated mode expansion becomes difficult.
Thus, we study the excitation of different frequency modes in this more complex system, including the emergence of Goldstone and Higgs modes, with the cavity output spectrum. Fig.~\ref{fig:11} shows the cavity output spectrum $S(\omega)$ for two different OAM configurations of the driving fields and for two different pump strengths. Figs. \ref{fig:11} (a)–(d) correspond to $(\ell_{1}, \ell_2) = (8,4)$, whereas Fig. \ref{fig:11} (e)–(h) correspond to the reversed configuration $(\ell_{1}, \ell_2) = (4,8)$. Since in this case the symmetry of scattered modes is broken, the excitation corresponding to $\ell_1$ and $\ell_2$ is individually probed by the spectra corresponding to modes $\alpha_+$ and $\beta_-$ respectively. In the weak pumping regime $\eta = 0.5\omega_{r}$, as shown in Figs. \ref{fig:11} (a), (b), (e), and (f), the spectrum exhibits only first-order peaks corresponding to $\omega_{\pm2}^{\ell_1}$ and $\omega_{\pm2}^{\ell_2}$ located at dashed vertical lines, which denote the analytically calculated frequency of first - order side modes as  
 \begin{equation}
        \omega_{\pm2}^{\ell_1} = \frac{\hbar}{2I}(L_p \pm 2\ell_1)^2
        =
        \begin{cases}
            \omega_{+2}^{\ell_1} = 8.03\,\omega_r, \\
            \omega_{-2}^{\ell_1} = 6.25\,\omega_r.
        \end{cases}
        \label{Eq:14}
 \end{equation}
 \begin{equation}
        \omega_{\pm2}^{\ell_2} = \frac{\hbar}{2I}(L_p \pm 2\ell_2)^2
        =
        \begin{cases}
            \omega_{+2}^{\ell_2} = 2.25\,\omega_r, \\
            \omega_{-2}^{\ell_2} = 1.36\,\omega_r.
        \end{cases}
        \label{Eq:15}
    \end{equation}
As the pump strength is increased to $100 \omega_r$, the cavity spectrum becomes significantly richer, showing the excitation of multiple collective modes in the system, show in in Figs. \ref{fig:11} (c), (d), (g), and (h). The lowest-frequency peak in the cavity spectrum disappears and reveals the emergence of a Goldstone mode associated with the continuous rotational symmetry breaking of the supersolid state, whereas the higher-frequency peaks, which correspond to amplitude (Higgs) modes shows shifts on increasing pump. These spectral signatures provide direct evidence of the collective excitation spectrum of the rotating supersolid phase in different $\ell_1$ and $\ell_2$ case and can be probed through the cavity output field.

{\subsection{Superposition of two rotational eigenstates}

Here, we consider the most general configuration, where the condensate occupies a coherent superposition of two rotational eigenstates $L_{p_1}$, $L_{p_2}$ and the pump fields carry asymmetric OAM $\ell_1\hbar$ and $-\ell_2\hbar$.

\subsubsection{Density Modulation}

To illustrate the resulting density structure, we focus on the representative case $(\ell_1, \ell_2) = (8, 4)$, with the condensate prepared in a superposition of the rotational states $L_{p_1}=2$ and $L_{p_2}=6$. The steady-state cavity-field amplitudes and the corresponding order parameters exhibit the same qualitative behavior as in the previously discussed asymmetric OAM configuration. As a result, the critical pump threshold remains unchanged.
The density modulation shows the $|L_{p_1}-L_{p_2}| = 4$ wavepackets with $2\ell_2 = 8$ number of fine stripes. The corresponding steady-state density profiles in real space are shown in Fig.~\ref{fig:DM_l12_Lp12}. Figs. \ref{fig:DM_l12_Lp12}(a) and \ref{fig:DM_l12_Lp12}(b) display the condensate density on the ring for two different pump strengths, $\eta=10\omega_r$ and $\eta=100\omega_r$, respectively. Below the threshold [Fig.~\ref{fig:DM_l12_Lp12}(a)], the density modulation appears as controlled by the superposition of $L_{p_1}$ and $L_{p_2}$, resulting in a pattern characterized by four prominent peaks superimposed on a weaker background. Above the threshold [Fig.~\ref{fig:DM_l12_Lp12}(b)], the modulation is controlled by pump OAM as well as superposition states. In this regime, the number of dominant peaks increases to 8, consistent with $2\ell_2$, and these peaks are confined within an overall wave packet envelope. Although the envelope is not clearly resolved in Fig.~\ref{fig:DM_l12_Lp12}(b), it becomes visible in Fig.~\ref{fig:DM_l12_Lp12}(c), where the modulation is shown at two different times, $t=0$ sec and $t=1$ sec. At the initial time, the modulation (red) is due to the interference of $L_{p_1}$ and $L_{p_2}$. At later time $t = 1$ sec, the modulation is a wavepacket with a number of packets $|L_{p_1} - L_{p_2}|$, and fine stripes of $2\ell_2 = 8$. Further insight is obtained from the angular-momentum distribution shown in Fig.~\ref{fig:DM_l12_Lp12}(d). The population is distributed over multiple neighbouring rotational states $L_{p_1} + n \ell_1$, $L_{p_1} + n \ell_1$, $L_{p_2} + n \ell_2$, and $L_{p_2} + n \ell_2$, where $n \in \{\pm2, \pm4......\}$.
\begin{figure}[t]
    \centering
    \includegraphics[width=\linewidth]{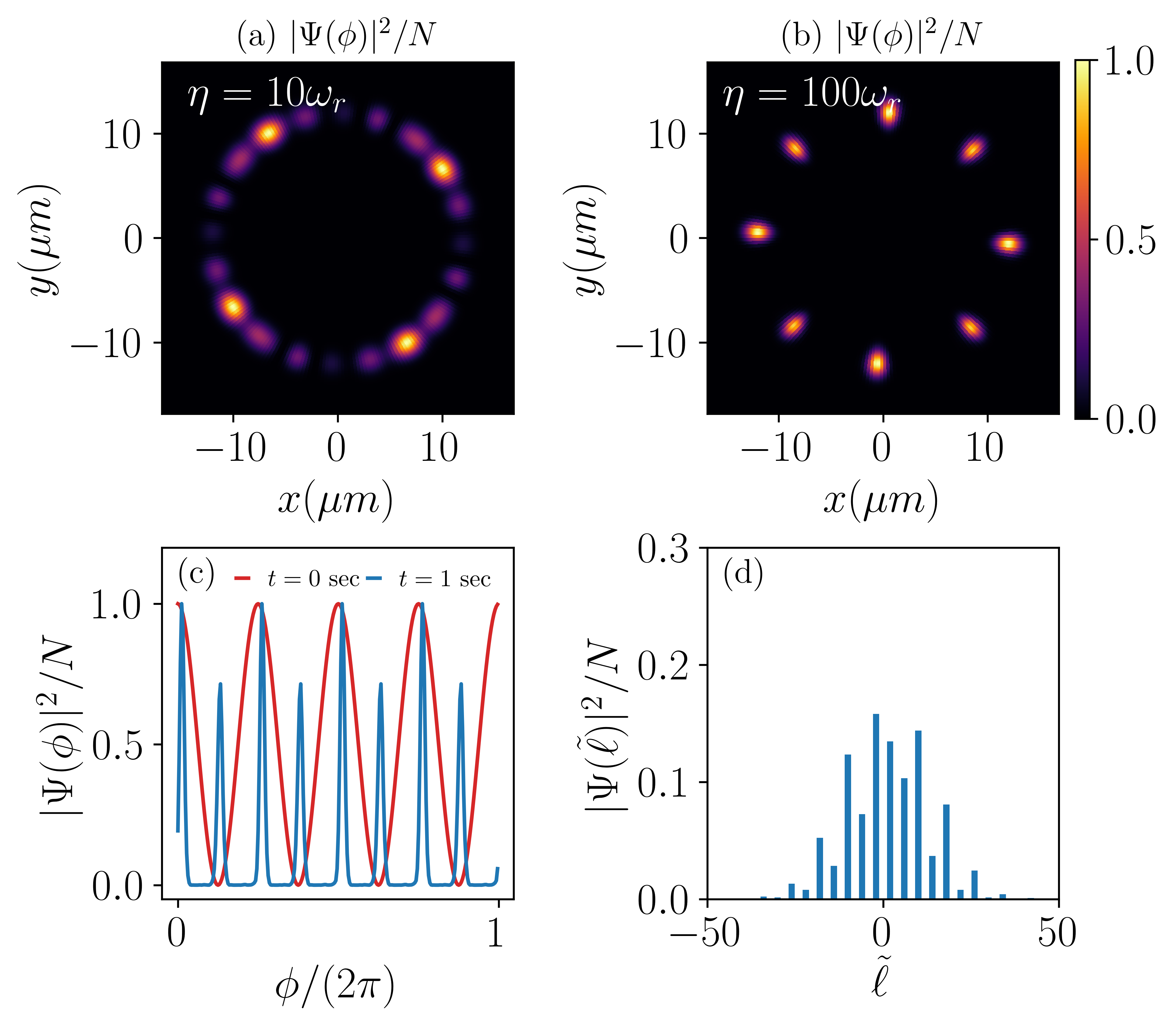}
    \caption{(a)-(b) The density modulation over the ring at $\eta = 10 \omega_r$, and $100 \omega_r$ respectively. (b) The density modulation as a function of $\phi$ at $t=0$ sec and $t=1$ sec.  (d) The Fourier transform of the BEC wavefunction shows the different momentum component excitation.}
    \label{fig:DM_l12_Lp12}
\end{figure}

The subsequent spatio-temporal evolution of density modulation further confirms the dynamical nature of the self-organized phase, shown in Fig. \ref{fig:l1_l2_Lz}(a). As time evolves, the localized density peaks shift uniformly along the azimuthal direction. Its average angular momentum and angular speed are shown in Fig. \ref{fig:l1_l2_Lz}(b), which shows a finite value and confirms the rotation with respect to time. The calculated steady state value of $\langle L_z \rangle /N \hbar = 3$, and $\Omega / 2 \pi N = 8$.

\begin{figure}[t]
    \centering
    \includegraphics[width=\linewidth]{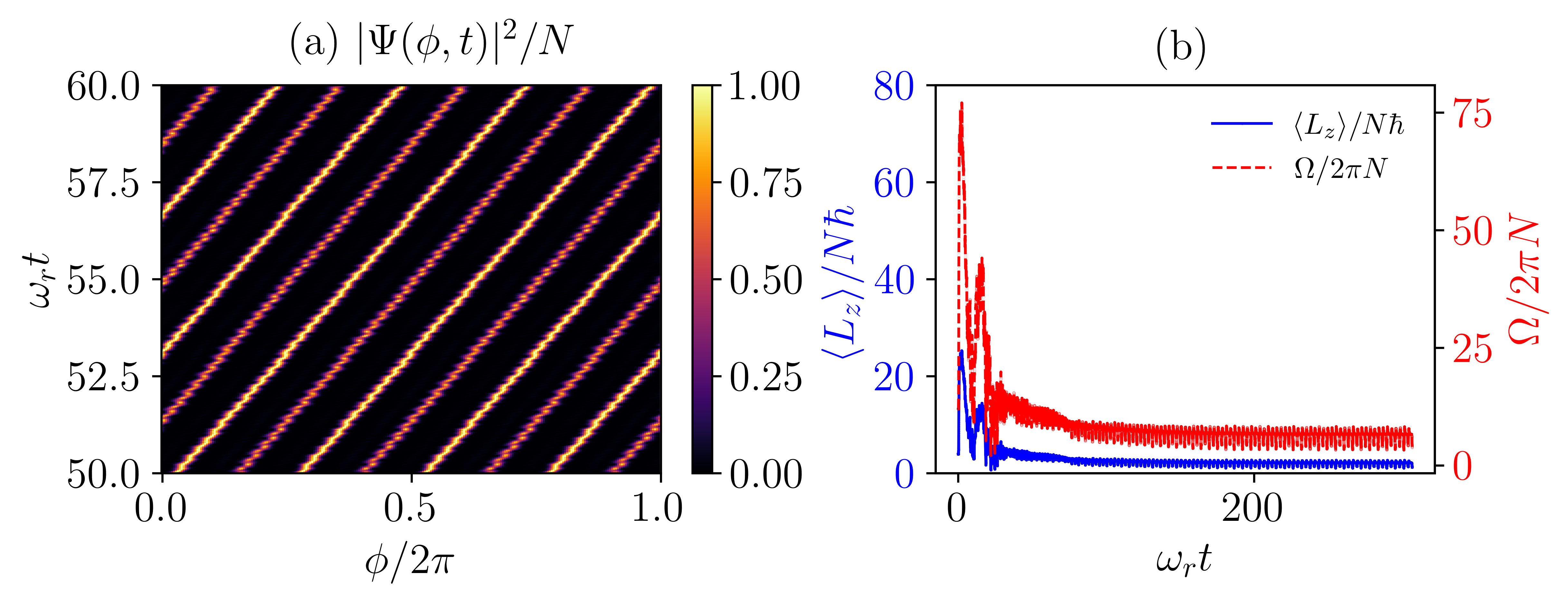}
    \caption{(a) The spatio-temporal evolution of density modulation. (b) The time dynamics of average angular momentum $\langle L_z \rangle / N \hbar$ (with blue), and angular velocity $\Omega/(2\pi N)$ (with red).}
    \label{fig:l1_l2_Lz}
\end{figure}
\section{Conclusion}
\label{SecV}
We have theoretically explored the formation of rotating supersolid phases in a ring BEC coupled to an optical ring cavity. Under symmetric LG beam driving, the system transitions from a homogeneous superfluid to a supersolid phase that simultaneously supports density modulation and global phase coherence. The reversibility of this transition under variations of the pump strength highlights its robustness. This behavior originates from the spontaneous breaking of the continuous rotational symmetry and is accompanied by the emergence of gapless Goldstone gapped Higgs modes, which provide clear signatures of supersolidity accessible via the cavity output.

When the condensate is initialized in a coherent superposition of angular-momentum eigenstates, a qualitatively distinct supersolid regime emerges. Instead of a uniform density modulation, the system develops localized, rotating wave packets.  Their number and internal structure are determined by the angular-momentum imbalance between the condensate initial state and the driving optical fields. The resulting rotation emerges intrinsically from coherent interference among angular-momentum modes, without requiring externally imposed rotation, i.e., such as optical \cite{PhysRevLett.110.025302} or magnetic \cite{PhysRevLett.124.025301,Pandey2019} stirring.

Further control is achieved under asymmetric optical driving with unequal orbital angular momenta. This introduces competing self-ordering channels and controllable chirality to the system, enabling independent manipulation of order parameters, the amplitudes of the scattered cavity modes, the spatial structure of the density modulation, and the direction of rotation. In  this regime, the condensates initialized in a single winding-number state, develop rotating density stripes whose angular periodicity is primarily set by the lower orbital angular momentum component of the driving field. When the condensate is prepared in a coherent superposition of winding number eigenstates, localized rotating wave packets emerge, with their separation set by the difference in winding numbers, while the finer stripe-like substructure within each packet is determined by the smaller OAM component of the optical drive. 

Overall, our system offers a highly tunable platform in which the rotational dynamics, spatial periodicity, and symmetry of the supersolid can be engineered through the orbital angular momentum of light \cite{PhysRevA.111.033304}. The coexistence of uniform and packetized supersolid phases, together with intrinsically generated rotation, highlights its versatility. Our results establish cavity-coupled ring condensates as a promising setting for exploring nonequilibrium supersolidity \cite{PhysRevLett.120.123601}, chiral quantum phases \cite{PhysRevLett.121.030404}, and angular-momentum–engineered quantum matter \cite{PhysRevLett.121.113204, PhysRevLett.122.045302}.

\section*{Acknowledgments}
M.B. thanks the Air Force Office of Scientific Research (AFOSR) (FA9550-23-1-0259) for support. We would like to thank Dr. Aritra Ghosh for useful discussions and a critical reading of the paper.

\appendix

\section{Effective Hamiltonian}
\label{sec:appendixA}
\noindent The ring cavity supports four traveling-wave optical modes $a_\pm, b_\pm$ with well-defined OAM. The optical field interacting with the BEC can be expressed in terms of these modes
\begin{equation}
\begin{aligned}
    E(\phi, t) = & E_0 u_{\ell_1}[a_+(t) e^{i \ell_1 \phi} + a_-(t) e^{-i \ell_1 \phi}]\hat{e}_a\\
    & + E_0u_{\ell_2}[b_+(t) e^{i \ell_2 \phi} + b_-(t) e^{-i \ell_2 \phi}]\hat{e}_b. \\
\end{aligned}
\end{equation}
Here, $\hat{e}_i$ denotes the polarization unit vector and $i=(a,b)$. The term $E_0$ is the field amplitude, and $u_{\ell_1}, u_{\ell_2}$ represent the normalized transverse mode profiles of the OAM mode corresponding to $\ell_1$ and $\ell_2$, respectively. A real electric field can be written as the sum of its positive and negative frequency parts
\begin{equation}
    E_T(\phi, t) = E^+(\phi, t)e^{-i\omega_l t} + E^-(\phi, t)e^{i\omega_l t},
\end{equation}
where, $E^-(\phi, t) = [E^+(\phi, t)]^\dagger$.
The dipole interaction Hamiltonian between the atom having the dipole moment vector $(\vec{d})$ and the total electric field $\vec{E}_T$ is given by
\begin{equation}
\begin{aligned}
    H_I = &  -\vec{d} \cdot \vec{E}_T(\phi, t),\\
    =  & -\left( \vec{d}_{eg} \sigma_+ + \vec{d}_{ge} \sigma_- \right) \cdot \left( E^+(\phi,t) + E^-(\phi,t) \right),
\end{aligned}
\end{equation}
where $\sigma_+ = |e\rangle \langle g|$ and $\sigma_- = |g\rangle \langle e|$ are the atomic raising and lowering operators. The coupling constant is defined by $g_0 = -{\vec{d}_{eg} \cdot \hat{e}_i E_0}/{\hbar} = -d\sqrt{{\omega_c}/{( \hbar \epsilon_0 V)}}$, where $V$ is the effective mode volume and $\omega_c$ the cavity resonance frequency. We considered $|\vec{d}_{eg}| = |\vec{d}_{ge}|$ implies $g_0 = g_0^*$ and both fields are interacting equally with the atom.  
The interaction Hamiltonian thus simplifies to
\begin{equation}
\begin{aligned}
H_I = & \hbar g_0 \sigma_+ \Big[
    (a_+ e^{i \ell_1 \phi} + a_- e^{-i \ell_1 \phi})e^{-i \omega_l t}\\
    & + (b_+ e^{i \ell_2 \phi} + b_- e^{-i \ell_2 \phi}) e^{-i \omega_l t}\Big] + \text{h.c.}
\end{aligned}
\end{equation}
To simplify the dynamics, we move to a frame rotating at the pump frequency $\omega_l$ using the unitary transformation: $U(t) = \exp\left[i\omega_l t\left(\sigma_+\sigma_- + a_+^\dagger a_+ + a_-^\dagger a_- + b_+^\dagger b_+ + b_-^\dagger b_-\right)\right].$  Applying the transformation and the rotating-wave approximation, the Hamiltonian, which includes the atomic contribution, atom–field interaction, and the driving term, can be expressed as
\begin{equation}
\begin{aligned}
    H_{\mathrm{rot}}   = & \frac{-\hbar^2}{2I}\frac{\partial^2}{\partial \phi ^2} - \hbar \Delta_a \sigma_+ \sigma_- + \hbar g_0 \Big[ \sigma_+ \big((a_+ e^{i \ell_1\phi}\\
    & + a_- e^{-i \ell_1\phi}) + (b_+ e^{i \ell_2\phi} + b_- e^{-i \ell_2\phi}) \big) + \mathrm{h.c.} \Big]\\
    & - \hbar \Delta_c \Big[ a_+^\dagger a_+ + a_-^\dagger a_- +  b_+^\dagger b_+ + b_-^\dagger b_-\Big]\\
    & - i\hbar\Big[ \eta_+ (a_+ - a_+^\dagger) + \eta_- (b_- - b_-^\dagger) \Big],
\end{aligned}
\end{equation}
where the detunings are defined by $
\Delta_a = \omega_l - \omega_a$, and $\Delta_0 = \omega_l - \omega_c.$
Next, in the regime of large detuning and low saturation, we adiabatically eliminate the atomic excited state, yielding effective cavity-field equations with atom-induced dispersive and dissipative terms. These are given, respectively,  by 
\begin{equation}
    U_0 = \frac{\Delta_a}{\Delta_a^2 + \gamma^2}g_0^2, \quad \Gamma_0 = \frac{\gamma}{\Delta_a^2 + \gamma^2}g_0^2\;.  
\end{equation}
Here, $U_0$ is the dispersive frequency shift per photon and, in the dispersive regime ($\Delta_a\gg\gamma$), takes the form $U_0 \approx g^2 /\Delta_a$. Meanwhile, $\Gamma_0$ denotes the incoherent scattering rate, which varies inversely with the square of atomic detuning in the same regime. In the dispersive limit, $U_0\gg\Gamma_0$ and the system is governed primarily by coherent interactions \cite{RevModPhys.85.553}.

In an atomic bosonic condensate of $N$ identical atoms, all atoms reside in the same quantum state. When two orthogonally polarized beams  are considered, the interaction terms between the 
$a_\pm$ and $b_\pm$ modes disappear due to destructive interference, and the resulting complete many-body Hamiltonian in the second-quantization picture takes the form presented in Eq. \eqref{Eq:1}.

\section{Symmetric OAM}
\label{sec:appendixB}

\subsection{Single winding number state}
\label{sec:appendixB1}
\subsubsection{Collective excitation matrix}
\label{sec:appendixB1a}
To analyze the collective excitation, we expand the cavity field and atomic operators around their mean-field values and retain terms up to linear order in the quantum fluctuations. Each operator is decomposed as $\mathcal{O}(t) = \mathcal{O}_{s} + \delta \mathcal{O}(t)$,
where $\mathcal{O}_{s}$ denotes the steady-state expectation value of the cavity-field operator and $\delta \mathcal{O}(t)$ represents small quantum fluctuations. Similarly, the atomic momentum amplitudes are expanded as $c_j(t) = e^{-i \mu_0 t / \hbar}\left[c_j + \delta c_j(t)\right]$,
with $\mu_0$ being the chemical potential of the condensate.
To investigate the role of quantum fluctuations in the vicinity of the instability threshold, we truncate the momentum eigenbasis to first order around the macroscopically occupied modes. The resulting linearized equations couple the fluctuations $\delta \mathcal{O}$ and $\delta c_j$ to their complex conjugates, reflecting the particle-hole mixing characteristic of driven-dissipative condensates. We therefore employ the Bogoliubov ansatz
\begin{align}
\delta \mathcal{O}(t) &= \delta \mathcal{O}^{(+)} e^{-i \omega t}
+ \delta \mathcal{O}^{(-)} e^{i \omega^{*} t}, \\
\delta c_j(t) &= \delta c_j^{(+)} e^{-i \omega t}
+ \delta c_j^{(-)} e^{i \omega^{*} t}.
\end{align}
Substituting this ansatz into the linearized equations of motion and independently equating the coefficients of $e^{-i \omega t}$ and $e^{i \omega^{*} t}$ to zero yields a set of coupled equations for the positive- and negative-frequency components of the fluctuations. These equations can be cast into a Bogoliubov-de Gennes (BdG) form. 
The resulting equations of motion can be written in a BdG form by the fluctuation vector
\begin{equation}
\begin{aligned}
    \mathbf{v} =  
    (\delta \alpha_+^{(\pm)},
     \delta \alpha_-^{(\pm)}, 
     \delta \beta_+^{(\pm)}, 
     \delta \beta_-^{(\pm)},
     \delta c_{+2}^{(\pm)},
     \delta c_{-2}^{(\pm)})^T.
\end{aligned}
\end{equation}
In this basis, the BdG matrix takes the block form
\begin{equation}
\mathcal{M}_{\mathrm{BdG}} =
\begin{pmatrix}
M_C & M_{CA} \\
M_{AC} & M_A
\end{pmatrix},
\end{equation}
where $M_C$ describes the cavity-field fluctuations, $M_A$ corresponds to the atomic sector, and $M_{CA}$ ($M_{AC}$) accounts for the cavity-atom (atom-cavity) coupling. The different block matrices are calculated as
\begin{align}
M_C &= 
\begin{pmatrix}
\Delta & 0 & \mathcal{D} & 0 & 0 & 0 & 0 & 0 \\
0 & -\Delta^* & 0 & -\mathcal{D}^* & 0 & 0 & 0 & 0 \\
\mathcal{D}^* & 0 & \Delta & 0 & 0 & 0 & 0 & 0 \\
0 & -\mathcal{D} & 0 & -\Delta^* & 0 & 0 & 0 & 0 \\
0 & 0 & 0 & 0 & \Delta & 0 & \mathcal{D} & 0 \\
0 & 0 & 0 & 0 & 0 & -\Delta^* & 0 & -\mathcal{D}^* \\
0 & 0 & 0 & 0 & \mathcal{D}^* & 0 & \Delta & 0 \\
0 & 0 & 0 & 0 & 0 & -\mathcal{D} & 0 & -\Delta^*
\end{pmatrix}\;,\\
M_{CA} &= U_0
\begin{pmatrix}
\alpha_- c_0^* & 0 & 0 & \alpha_- c_0 \\
0 & -\alpha_-^* c_0 & -\alpha_-^* c_0^* & 0 \\
0 & \alpha_+ c_0 & \alpha_+ c_0^* & 0 \\
-\alpha_+^* c_0^* & 0 & 0 & -\alpha_+^* c_0 \\
\beta_- c_0^* & 0 & 0 & \beta_- c_0 \\
0 & -\beta_-^* c_0 & -\beta_-^* c_0^* & 0 \\
0 & \beta_+ c_0 & \beta_+ c_0^* & 0 \\
-\beta_+^* c_0^* & 0 & 0 & -\beta_+^* c_0
\end{pmatrix}\;,\\
M_A &= \text{diag}\Big(M_{+}, -M_{+}, M_{-}, -M_{-} \Big)\;,
\end{align}
\noindent where we introduce $\Delta = -(\delta_c + i \kappa)$, $\mathcal{D}=U_{0}\mathcal{N}$, $\mathcal{N} = (c_0^* c_{+2} + c_{-2}^* c_0)$, and $M_{+(-)}=(\omega_{+(-)2} - \mu/\hbar)$. The reverse coupling block satisfies $M_{AC} = M_{CA}^\dagger $. 
The reverse coupling block satisfies $M_{AC} = M_{CA}^\dagger $. Here, $\mu$ is the chemical potential, determined by using $c_j(t) = c_j e^{-i\mu t}/\hbar$ and applying the equation of motion for the $c_j$ modes.
\begin{equation}
\begin{aligned}
        \mu c_0/\hbar = &  \omega_0 c_0 + U_0 \left[\mathcal{A}c_{+2} + \mathcal{A}^* c_{-2} \right],\\
        \mu c_{+2}/\hbar = & \omega_{+2} c_{+2} + U_0 \left[\mathcal{A}^*c_0  \right],\\ 
        \quad \mu c_{-2}/\hbar = & \omega_{-2} c_{-2} + U_0 \left[\mathcal{A} c_0  \right],
\end{aligned}
\end{equation}
where, $\mathcal{A} = \alpha_+^* \alpha_- + \beta_+^* \beta_-$. Combining these three equations, we get 
\begin{equation}
    \mu/\hbar = \omega_0 |c_0|^2 + \omega_+|c_{+2}|^2 + \omega_-|c_{-2}|^2 + U_0 \left[\mathcal{A}\mathcal{N} + \mathcal{A}^* \mathcal{N}^* \right]
\end{equation}
The excitation spectrum is obtained by diagonalizing the BdG matrix $\mathcal{M}_{\mathrm{BdG}}$.

\subsubsection{Cavity output spectrum}
\label{sec:appendixB1b}
\begin{figure*}[htbp]
    \centering
    \includegraphics[width=\linewidth]{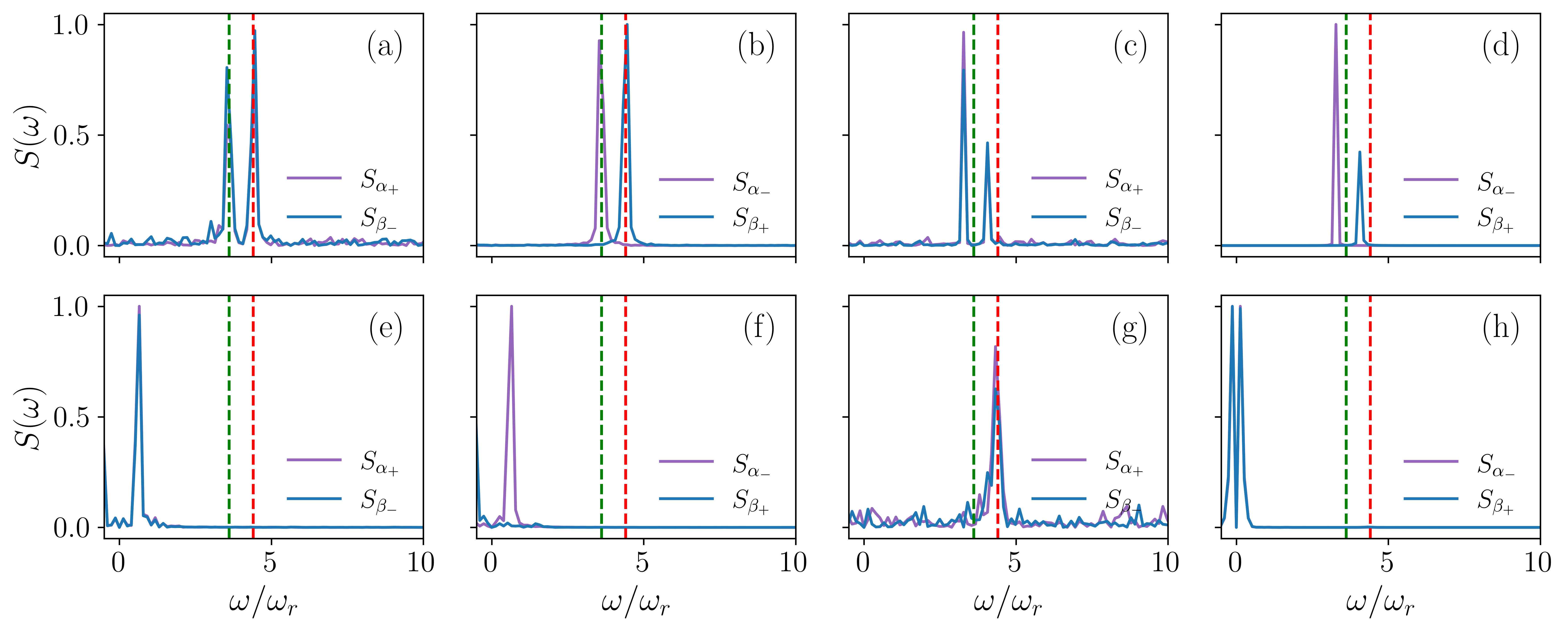}
    \caption{Cavity output spectra of the four cavity modes for a single winding number state $L_p$ at different pump strengths $\eta/\omega_r$. (a)–(b) $\eta = 0.5\omega_r$, (c)–(d) $\eta = 10\omega_r$, (e)–(f) $\eta = 20\omega_r$, and (g)–(h) $\eta = 30\omega_r$. The red and green dotted lines indicate the analytical values of the frequencies $\omega_{\pm2}$ as given in Eq. \eqref{Eq:11}.}
    \label{fig:14}
\end{figure*}
The cavity output spectra corresponding to all four cavity modes are shown in Fig.~\ref{fig:14}. The spectra are plotted for different pump strengths as specified in the figure caption. The spectra associated with the pump modes $S_{\alpha_+}(\omega)$ and $S_{\beta_-}(\omega)$ exhibit the presence of the characteristic frequencies $\omega_{\pm 2}$.
The scattered-field spectra resolve the individual excitations more distinctly. Specifically, the spectrum $S_{\alpha_-}(\omega)$ predominantly shows the $\omega_{-2}$ mode, while $S_{\beta_+}(\omega)$ reveals the $\omega_{+2}$ mode. The observed peak positions shift with pump strength and follow the expected Bogoliubov dispersion.
These results indicate that monitoring the spectrum of either of the pump modes alone is sufficient to extract information about both the Goldstone and Higgs excitations of the system. The cavity output thus provides a direct probe of the collective excitation spectrum.

\subsection{Superposition of two rotational eigenstates}
\label{sec:appendix:B2}

\subsubsection{Density modulation}
\label{sec:appendixB2a}
The different sets of $L_{p_1}$ and $L_{p_2}$, show the various density profiles, with the number of packets equal to $|L_{p_1} - L_{p_2}|$ as discussed in the main text and rotating with constant velocity over the ring. For the case $L_{p_1} = 1$ and $L_{p_2} = -1$, the lattice is purely static with zero velocity, because of the destructive interference between $L_{p_1}$ and $L_{p_2}$.
The comparison between these two cases further illustrates how the choice of initial winding numbers controls the number of packets and the rotation dynamics of the supersolid.
\begin{figure}[b]
    \centering
    \includegraphics[width=\linewidth]{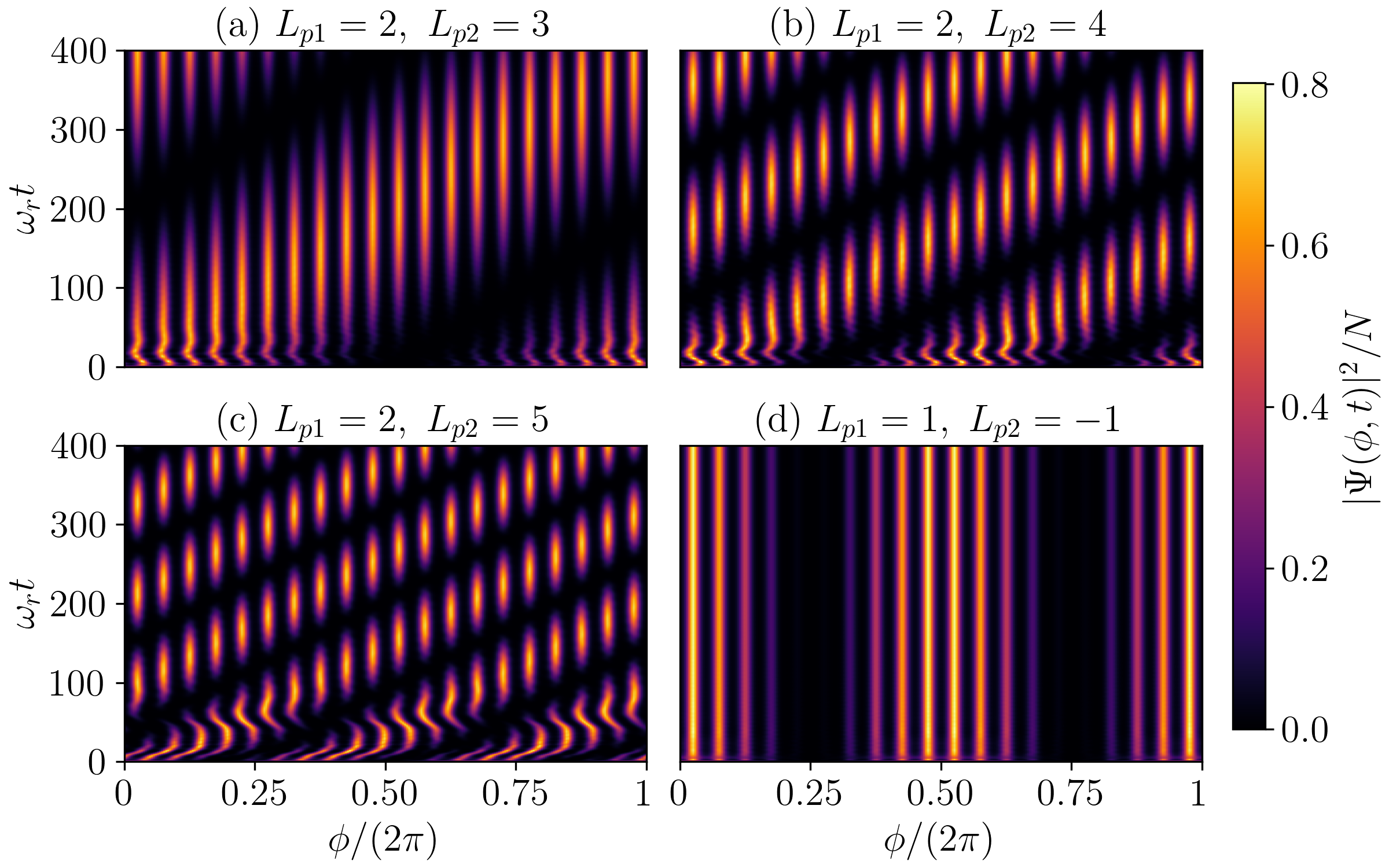}
    \caption{Spatiotemporal modulation of BEC for different combinations of $L_{p_1}$ and $L_{p_2}$. (a) $L_{p_1} = 2, L_{p_2} = 3$, (b) $L_{p_1} = 2, L_{p_2} = 4$, (c) $L_{p_1} = 2, L_{p_2} = 5$, (a) $L_{p_1} = 1, L_{p_2} = -1$.}
    \label{fig:15}
\end{figure}
\noindent Considering the initial state of BEC in the equal and symmetric superposition of winding numbers $L_{p_1}$ and $L_{p_2}$. The equations of motion can then be solved and the resulting time dynamics of density modulation w.r.t $\phi$ for different combinations of $L_{p_1}$ and $L_{p_2}$ are shown in Fig. \ref{fig:15}.
\subsubsection{Collective excitation matrix}
\label{sec:appendixB2b}
\noindent Again, we use the same approach as discussed in single $L_p$ case and defines the fluctuation vector as
\begin{equation}
\begin{aligned}
\mathbf{v} = 
(\delta \alpha_+^{(\pm)},
 \delta \alpha_-^{(\pm)},
 \delta \beta_+^{(\pm)}, 
 & \delta \beta_-^{(\pm)}, 
 \delta c_{+2}^{(\pm)}, 
 \delta c_{-2}^{(\pm)}, 
 \delta d_{+2}^{(\pm)}, \\
 & \delta d_{-2}^{(\pm)})^T .
 \end{aligned}
\end{equation}
The matrix corresponding to cavity-field fluctuation $M_C$ is the same as that written in the single $L_p$ case, The only modification arises in the definition of the coupling term, where
\begin{equation}
    \mathcal{N} = c_0^* c_{+2} + c_{-2}^* c_0 + d_0^* d_{+2} + d_{-2}^* d_0,
\end{equation}
is replaced by the appropriate generalized expression for the present configuration.
To determine the chemical potential, we assume a time-dependent form of the atomic amplitudes as $c_j(t) = c_j e^{-i \mu t/\hbar}$. Substituting this ansatz into the corresponding equations of motion and consistently combining them, we obtain an explicit expression for the chemical potential $\mu$.
Using this result, the atomic matrix as well as the cavity–atom coupling matrix can be derived, which are presented below.
\begin{widetext}
\begin{equation}
\begin{aligned}
            M_A = & \text{diag}\Big((\omega_{+2}^{p_1} - \mu/\hbar), -(\omega_{+2}^{p_1} - \mu/\hbar), (\omega_{-2}^{p_1} -  \mu/\hbar),
             -(\omega_{-2}^{p_1} - \mu/\hbar), 
            (\omega_{+2}^{p_2} - \mu/\hbar),
            -(\omega_{+2}^{p_2} - \mu/\hbar),\\
            & (\omega_{-2}^{p_2} - \mu/\hbar), -(\omega_{-2}^{p_2} - \mu/\hbar) \Big),
\end{aligned}
\end{equation}
\begin{equation}
\begin{aligned}
    \mu/\hbar =  \omega_0^{p_1}|c_0|^2 + \omega_0^{p_2}|d_0|^2 + \omega_{+2}^{p_1}|c_{+2}|^2 
     + \omega_{-2}^{p_1}|c_{-2}|^2 + \omega_{+2}^{p_2}|d_{+2}|^2 + \omega_{-2}^{p_2}|d_{-2}|^2
     + U_0 \left[\mathcal{A}\mathcal{N} + \mathcal{A}^* \mathcal{N}^* \right]\;,
\end{aligned}
\end{equation}
\begin{equation}
    M_{CA} = U_0 \begin{pmatrix}
        \alpha_- c_0^* & 0 & 0 & \alpha_- c_0 & \alpha_- d_0^* & 0 & 0 & \alpha_- d_0\\
        0 & -\alpha_-^* c_0 & -\alpha_-^* c_0^* & 0 & 0 & -\alpha_-^* d_0 & \alpha_-^* d_0^* & 0\\
        0 & \alpha_+ c_0 & \alpha_+ c_0^* & 0 & 0 & \alpha_+ d_0 & \alpha_+ d_0^* & 0\\
        -\alpha_+^* c_0^* & 0 & 0 & -\alpha_+^* c_0 & \alpha_+^* d_0^* & 0 & 0 & -\alpha_+^* d_0\\
        \beta_- c_0^* & 0 & 0 & \beta_- c_0 & \beta_- d_0^* & 0 & 0 & \beta_- d_0\\
        0 & -\beta_-^* c_0 & -\beta_- c_0^* & 0 & 0 & -\beta_-^* d_0 & \beta_-^* d_0^* & 0\\
        0 & \beta_+ c_0 & \beta_+ c_0^* & 0 & 0 & \beta_+ d_0 & \beta_+ d_0^* & 0\\
        -\beta_+^* c_0^* & 0 & 0 & -\beta_+^* c_0 & \beta_+^* d_0^* & 0 & 0 & -\beta_+^* d_0.
    \end{pmatrix}.
\end{equation}
The reverse coupling block satisfies $M_{AC} = M_{CA}^\dagger $.
\end{widetext}
%

%

\end{document}